\documentclass[twocolumn,tighten]{aastex61}
\pdfoutput=1 %for arXiv submission
\usepackage{amsmath,amstext}
\usepackage{apjfonts} 
\usepackage{afterpage}
\usepackage{xcolor}

\newcommand{\march}{\texttt{MARCH}}

\usepackage{graphicx}
\usepackage{epsfig} 

\usepackage[normalem]{ulem}

\shorttitle{\march}
\shortauthors{Psaltis et al.}

\begin{document}


\title{Markov Chains for Horizons (\march).\ I.\  Identifying Biases in Fitting Theoretical Models to Event Horizon Telescope Observations}
\author{Dimitrios Psaltis}
\affiliation{Department of Astronomy, University of Arizona, 933 N. Cherry Ave, Tucson, AZ 85721, USA}

\author{Feryal \"Ozel}
\affiliation{Department of Astronomy, University of Arizona, 933 N. Cherry Ave, Tucson, AZ 85721, USA}

\author{Lia Medeiros}
\affiliation{Department of Astronomy, University of Arizona, 933 N. Cherry Ave, Tucson, AZ 85721, USA}
\affiliation{School of Natural Sciences, Institute for Advanced Study, 1 Einstein Drive, Princeton, NJ 08540, USA}

\author{Pierre Christian}
\affiliation{Department of Astronomy, University of Arizona, 933 N. Cherry Ave, Tucson, AZ 85721, USA}

\author{Junhan Kim}
\affiliation{Department of Astronomy, University of Arizona, 933 N. Cherry Ave, Tucson, AZ 85721, USA}
\affiliation{California Institute of Technology, 1200 E. California Blvd., MC 367-17, Pasadena, CA 91125, USA}

\author{Chi-kwan Chan}
\affiliation{Department of Astronomy, University of Arizona, 933 N. Cherry Ave, Tucson, AZ 85721, USA}

\author{Landen J.\ Conway}
\affiliation{Department of Astronomy, University of Arizona, 933 N. Cherry Ave, Tucson, AZ 85721, USA}

\author{Carolyn A.\ Raithel}
\affiliation{Department of Astronomy, University of Arizona, 933 N. Cherry Ave, Tucson, AZ 85721, USA}

\author{Dan Marrone}
\affiliation{Department of Astronomy, University of Arizona, 933 N. Cherry Ave, Tucson, AZ 85721, USA}

\author{Tod R.\ Lauer}
\affiliation{National Optical Astronomy Observatory, 950 North Cherry Ave., Tucson, AZ 85719, USA}

\begin{abstract}
We introduce a new Markov Chain Monte Carlo (MCMC) algorithm with parallel tempering for fitting theoretical models of horizon-scale images of black holes to the interferometric data from the Event Horizon Telescope (EHT). The algorithm implements forms of the noise distribution in the data that are accurate for all signal-to-noise ratios. In addition to being trivially parallelizable, the algorithm is optimized for high performance, achieving 1 million MCMC chain steps in under 20 seconds on a single processor. We use synthetic data for the 2017 EHT coverage of M87 that are generated based on analytic as well as General Relativistic Magnetohydrodynamic (GRMHD) model images to explore several potential sources of biases in fitting models to sparse interferometric data. We demonstrate that a very small number of data points that lie near salient features of the interferometric data exert disproportionate influence on the inferred model parameters. We also show that the preferred orientations of the EHT baselines introduce significant biases in the inference of the orientation of the model images. Finally, we discuss strategies that help identify the presence and severity of such biases in realistic applications.
\end{abstract}

\section{INTRODUCTION}

Images of low-luminosity black holes at millimeter wavelengths, where the accretion flow is transparent down to horizon scales~\citep{Ozel2000}, have long been expected to be compact, dominated by the shadow cast by the black hole on the surrounding plasma emission~\citep{Falcke2000,Dexter2009,Broderick2009,Moscibrodzka2009,Moscibrodzka2014,Chan2015}. Recent Event Horizon Telescope (EHT) observations of the black hole in the center of the M87 galaxy have confirmed this expectation, revealing an image consisting of a narrow ring of emission surrounding the black-hole shadow~\citep{PaperI,PaperII,PaperIII,PaperIV,PaperV,PaperVI}.

As a Very Long Baseline Interferometric (VLBI) array, the EHT measures the complex Fourier components of the brightness distribution in the sky, also called the interferometric visibilities, at a distinct set of spatial frequencies determined by the baselines between the various stations in the array. In principle, the image of the source can then be obtained by constructing the inverse Fourier transform of the observed visibilities. In practice, however, the sparse coverage of Fourier space by the EHT, which consists of a small number of baselines separated by large distances, makes such a direct reconstruction of the image impossible.

Traditionally, VLBI measurements have been converted to images with CLEAN algorithms (see, e.g., \citealt{Hogbom1974,Clark1980}), which identify the minimum set of point sources in the image plane with Fourier transforms that are consistent with the visibility data. These point sources are then convolved with a filter to broaden the points sources to the effective resolution of the interferometric array. 

Alternatively, interferometric images have been obtained with regularized maximum likelihood methods in which a pixel-based model is defined in the image plane with its parameters being the brightness at each pixel location. The Fourier transform of this model is compared to the observed visibilities and the brightness of each pixel is adjusted to minimize an appropriate $\chi^2$-statistic in visibility space. Because the number of pixels of the model image is typically much larger than the number of measurements, regularizers are employed to decrease the effective number of degrees of freedom.  This approach was introduced in radio interferometry imaging with maximum-entropy regularizers (see~\citealt{Narayan1986} and references therein), has been developed extensively in optical/IR interferometry~\citep{Thiebaut2009,Berger2012}, and has been employed in imaging black holes with the EHT (see, e.g.,~\citealt{Honma2014,Chael2016,Chael2018,Akiyama2017, Akiyama2018,PaperIV}).

The regularized maximum likelihood methods are fundamentally non-parametric model fitting algorithms. As such, they are agnostic in regards to the underlying image structure. However, this freedom comes with the usual ``curse of dimensionality'': the requirement of quantifying the brightness on a very large number of pixels typically leads to presenting a single (or a very small number of) best-fit images with marginal exploration of the parameter space of possible solutions that might fit the data equally well. Moreover, the image properties show a dependence on the regularizers and the strength of the regularizers is usually chosen such that they suppress structures at scales smaller than the effective beam of the array~(see, e.g., \citealt{PaperIV}). This limits the ability of the algorithms to provide precise measurements of physical quantities of interest. 

In order to elucidate the limitations of an agnostic imaging algorithm, we can consider interferometric observations of a simple source consisting of two point sources separated by some angular distance. An imaging algorithm will only be able to measure the separation of the two point sources down to an accuracy comparable to the effective beam size of the array, which depends mainly on the wavelength of observation and the separations of the interferometric stations. However, if we know a priori that we are looking at an image with two point sources, we can use the same interferometric data to measure the angular separation of the point source with a much higher accuracy that will be limited primarily by the signal-to-noise ratio of the measurements. Achieving such a measurement for a realistic image, of course, requires both an accurate parametric model for the underlying image as well as an algorithm to explore the posterior distribution of the statistical comparison of the model to the data.

In the first article of this series, we describe the new Markov-Chain Bayesian algorithm \march\ (MArkov Chains for Horizons) that we have developed to fit models of black-hole images to the interferometric data at mm wavelengths obtained with the EHT. Similar Bayesian approaches have already been applied to early EHT observations of the galactic center black hole Sgr~A* (see, e.g.,~\citealt{Broderick2016,Lu2018}) as well as to the 2017 EHT observations of M87 that have led to the initial measurements of the mass of the latter black hole~\citep{PaperVI}. Our algorithm aims to address a number of complications to this otherwise straightforward statistical problem introduced by theoretical expectations on the images of black holes and the particular characteristics of the EHT array.

First, the structure of the mm image of Sgr~A* is expected to show substantial variability within the few-hour time span of the observations because of its small mass~(see, e.g.,~\citealt{Medeiros2017, Medeiros2018, Roelofs2017}). As a result, fitting a static model image to variable interferometric data will introduce biases to the inferred parameters and artificially reduce their inferred uncertainties. This complication was addressed in earlier work~\citep{Kim2016} and we will not revisit it here. 

Second, the images of black holes with horizon-scale resolution are expected to have ring-like structures with very sharp boundaries at the locations of the black-hole shadows~(see, e.g.,~\citealt{Kamruddin2013,Psaltis2015}). The interferometric visibilities of these structures show a number of salient features such as deep minima of the visibility amplitudes at particular baseline lengths, with rapid swings of the visibility phases across them. The location of the first minimum depends very strongly on the size of the ring of emission and the location of the second minimum depends on its fractional width~\citep{PaperVI}. While the strong dependence of the salient features on model parameters allows us to measure them accurately, the most likely values of the model parameters and their uncertainties depend disproportionately on how well the model describes the very small number of data points near the salient features. As we will show in \S3, the combination of the simplicity of the model images with the high signal-to-noise ratio of the measurements causes a straightforward application of Bayesian posteriors to return biased measurements of the model parameters.

Finally, the location of landmass and high mountains on the globe forces the EHT baselines to lie primarily along two orientations, N-S and E-W~\citep{PaperII}. As a result of the projection-slice theorem, the measured interferometric visibilities are primarily sensitive only to the projections of the underlying image along these two orientations. We will show in \S4 that this biases the inferred orientations of the model images to line up with the major orientations of the baselines. The origin of this bias is similar to the development of spurious ``knots'' along the baseline orientations in the reconstruction of otherwise smooth images from synthetic EHT data~(see Fig.10 of~\citealt{PaperIV}). Additional evidence for this effect comes from the initial application of geometric and GRMHD model fitting to the 2017 EHT data of M87, which revealed that the most likely black-hole orientation appears aligned with the orientations of the EHT baselines (see Fig.~6 of~\citealt{PaperVI}) and away from the orientation of the long-wavelength jet that emerges from the central black hole~(see Fig.~9 of~\citealt{PaperV}).

In \S2 below, we introduce the basic formalism of our Bayesian Markov Chain algorithm and its implementation. In \S3-5, we apply this algorithm to various cases of simple synthetic data chosen to explore the issues discussed above and provide strategies for their mitigation. In \S6, we explore realistic EHT data based on GRMHD simulations of M87 and, in \S7, we conclude by discussing our results and providing an outlook on the accuracy of measurements with the EHT. In subsequent papers in this series, we will discuss our implementation of modeling the unknown gains of the various interferometric stations and the biases introduced by this approach; we will also explore different information criteria that allow us to quantify the necessity of increasing the complexity of models in order to fit realistic data.

\section{The \march\ Algorithm}

\subsection{Interferometric Observables}

The goal of the \march\ algorithm is to estimate the posterior distributions over a number of parameters of a model by comparing to EHT observations the model image brightness in the sky $I(x,y;\vec{\theta})$; here $x$ and $y$ are angular coordinates and $\vec{\theta}$ is the vector of model parameters. As an interferometer, the EHT does not directly observe the brightness distribution of the sky but rather the complex visibilities defined by the van Cittert-Zernike theorem as
\begin{equation}
{\cal V}(u,v;\vec{\theta})\equiv\int\int e^{-2\pi(x u +y v)}I(x,y;\vec{\theta})\; dx dy\;.
\label{eq:vCZtheorem}
\end{equation}
In this equation, $u$ and $v$ are equal to the two components of the baseline vectors $\vec{b}$ that connect every pair of sites in the array, projected orthogonally to the line of sight to a particular source, and divided by the wavelength $\lambda$ of observation, i.e., $(u,v)\equiv\vec{b}/\lambda$. The brightness distribution in the image domain and the visibilities in the $u-v$ domain are clearly connected via a 2-dimensional Fourier transform.

For a general, non symmetric brightness distribution, the interferometric visibilities are complex numbers that can be described either in terms of their real and imaginary components or in terms of their amplitudes and phases. The former description is useful in quantifying the thermal noise in the measurements, which can be well approximated by independent Gaussian distributions in the real and imaginary components of the visibilities (see~\citealt{TMS2017}). However, because of atmospheric and instrumental effects, seldom do VLBI measurements allow for a precise determination of the complex visibilities. In practice, we write the measured visibility between the $i-$th and $j-th$ station as
\begin{equation}
V_{ij}\equiv V(u_{ij},v_{ij})=g_i g_j^* {\cal V}(u_{ij},v_{ij})\;,
\end{equation}
where $g_i$ and $g_j$ are complex ``gains'' of the two telescopes and a star superscript denotes complex conjugates. Hereafter, as in this equation, we will denote the visibilities between the two stations simply as $V_{ij}$ and suppress for brevity the argument of this quantity.

The magnitudes of the gain factors for the various telescopes can often be constrained with {\em a priori\/} calibration to within $\sim 10$\% levels~\citep{PaperIII}. On the other hand, the complex phases of the gains can fluctuate wildly and over very short timescales because of atmospheric instabilities and instrumental effects. To overcome this problem, VLBI measurements quote the amplitudes $\vert V_{ij}\vert$ of the complex visibilities between pairs of stations and the arguments of the complex bispectra (also known as closure phases) along a closed triangle of stations~\citep{Jennison1958,TMS2017}.
The latter is simply
\begin{eqnarray}
{\rm Arg}[V_{ij} V_{jk} V_{ki} ]&=&{\rm Arg}[g_i g_j^* g_j g_k^* g_k g_i^* {\cal V}_{ij} {\cal V}_{jk} {\cal V}_{ki} ]
\nonumber\\
&=& 
{\rm Arg}[\vert g_i\vert^2 \vert g_j\vert^2 \vert g_k\vert^2] + {\rm Arg}[ {\cal V}_{ij} {\cal V}_{jk} {\cal V}_{ki} ]\nonumber\\
&=& 
 {\rm Arg}[{\cal V}_{ij} {\cal V}_{jk} {\cal V}_{ki} ]\;,
\end{eqnarray}
i.e., the closure phases are not affected by the unknown phases of the telescope gains.

We pay three penalties by using visibility amplitudes and closure phases instead of the real and imaginary components of the complex interferometric visibilities. First, by combining three complex visibilities to form one closure quantity, we lose some of the available information. Second, the transformation between the brightness in the image domain and the observables in the visibility domain is no longer linear. As a result, adding a perturbation to an image that would only introduce minor variations in the complex visibilities could still lead to order unity changes in visibility amplitudes and closure phases at various baseline lengths. This is important in understanding the consequences of the strong dependence of model parameters on the salient features of the visibilities. Third, the error distributions of the visibility amplitudes and closure phases are not Gaussian, even for purely thermal noise. 

The likelihood of making a visibility amplitude measurement $V_{ij}$ given a true visibility amplitude of $V_{ij,0}$  is the Rice distribution~\citep{TMS2017}
\begin{equation}
{\cal L}_A(V_{ij};V_{ij,0},\sigma_{ij})=\frac{V_{ij}}{\sigma_{ij}^2}
\exp\left[-\frac{V_{ij}^2+V_{ij,0}^2}{2\sigma_{ij}^2}\right]
I_0\left(\frac{V_{ij}V_{ij,0}}{\sigma_{ij}^2}\right)\;,
\label{eq:Rice}
\end{equation}
where $I_0(x)$ is the modified Bessel function of zeroth order and $\sigma_{ij}$ is the dispersion of thermal noise introduced to the real and imaginary components of the visibility by the array configuration used. At large signal-to-noise ratios $V_{ij,0}/\sigma_{ij}$, the Rice distribution reduces to a Gaussian with the same dispersion. The computational cost of using the Rice distribution is only marginally higher than using the Gaussian approximation. 

The likelihood of making a closure phase measurement $\Phi_{ijk}$ is given by the triple convolution
\begin{eqnarray}
&&{\cal L}_\Phi(\Phi_{ijk};\Phi_{ijk,0},\sigma_{ij},\sigma_{jk},\sigma_{ki})=\int d\theta \int d\phi
L_{\rm ph}(\theta;{\rm SNR}_{ij}) \nonumber\\
&&\qquad L_{\rm ph}(\phi-\theta;{\rm SNR}_{jk})
L_{\rm ph}(\Phi_{ijk}-\Phi_{ijk,0}-\phi;{\rm SNR}_{ki})\;.
\label{eq:clos_like}
\end{eqnarray}
Here $\Phi_{ijk,0}$ is the true closure phase and $L_{\rm ph}(\theta;{\rm SNR}_{ij})$ is the likelihood of measuring for the phase of the $ij-$baseline an angle $\theta$ away from the true value. This last likelihood is given by~\citep{TMS2017}
\begin{eqnarray}
L_{\rm ph}(\theta;{\rm SNR}_{ij})&=&
\frac{e^{-\frac{{\rm SNR}_{ij}}{2}}}{4\pi \sigma_{ij}}
\left\{\sqrt{2\pi}~{\rm SNR}_{ij} \cos\theta e^{\frac{{\rm SNR}_{ij}^2 \cos^2\theta}{2}}\right.\nonumber\\
&&\qquad \times \left.\left[{\rm Erf}\left(\frac{{\rm SNR}_{ij} \cos\theta}{\sqrt{2}}\right)+1\right]+2\right\}\;,
\end{eqnarray}
where ${\rm SNR}_{ij}\equiv V_{ij,0}/\sigma_{ij}$ is the signal-to-noise ratio of the measurement. The triple convolution necessary to calculate the likelihood for the closure phase is too expensive computationally. \citet{Christian2019} developed an efficient numerical algorithm that allows us to obtain an accurate approximation to the complete likelihood~(\ref{eq:clos_like}), which we use here.
At the limit of large signal-to-noise ratios, this distribution can be well approximated by a Gaussian with dispersion
\begin{equation}
\sigma_{\Phi_{ijk}}=\left(
\frac{\sigma_{ij}^2}{V_{ij}^2}+
\frac{\sigma_{jk}^2}{V_{jk}^2}+
\frac{\sigma_{ki}^2}{V_{ki}^2}\right)^{1/2}\;.
\end{equation}

\subsection{Bayesian Posteriors}

Given a set of interferometric data points, we calculate the posterior distribution of the model parameters $\vec{\theta}$ using Bayes' theorem, i.e.,
\begin{equation}
P(\vec{\theta}\vert{\rm data})=C\; P_{\rm pri}(\vec{\theta})\; {\cal L}({\rm data}|\vec{\theta})\;.
\label{eq:bayes}
\end{equation}
Here, $P_{\rm pri}(\vec{\theta})$ is the prior distribution over the model parameters, ${\cal L}({\rm data}|\vec{\theta})$ is the likelihood that the set of observed data can be obtained from a given set of model parameters, and $C$ is a normalization constant.

The likelihood function is obtained by a proper combination of the likelihoods of the visibility amplitude and closure phase data, equations~(\ref{eq:Rice}) and (\ref{eq:clos_like}), respectively. Writing schematically the likelihood of the $i-$th data point as ${\cal L}_i({\rm data}\vert \vec{\theta})$ and assuming that all likelihoods are independent of each other (see~\citealt{Blackburn2019}), we obtain
\begin{equation}
{\cal L}({\rm data}|\vec{\theta})= \prod_i {\cal L}_i({\rm data}|\vec{\theta})
\label{eq:like}
\end{equation}

The priors over the model parameters depend on the specifics of the model considered. For the analytic crescent model we will be using below, we set all the priors to boxcar functions with limits that are much larger than the widths of the expected posteriors. For reasons that will be discussed in the following sections, the posteriors have very small fractional widths (often less than 1\%) and, therefore, the precise function form of such non-informative priors does not affect the final results.

\subsection{Markov Chains, Parallel Tempering, and Performance}

The sparse coverage of the $u-v$ plane with the EHT baselines results in a small number of independent measurements and hence warrants fitting models to the data that only have a relatively small number of  parameters. Moreover, the salient features of the interferometric data, such as the baseline lengths that correspond to minima in the visibility amplitude, allow for relatively accurate initial guesses for the values of several model parameters (see the discussion in \S3 and \S4 below). For these reasons, we employ a simple Metropolis-Hastings stepping algorithm for each chain of the MCMC algorithm. In particular, we choose steps for each model parameters from independent Gaussian distributions with dispersions set equal to a fraction of the expected uncertainties in the corresponding model parameters.

The presence of salient features in the interferometric data often leads to narrow widths for the posteriors and strong correlations between some model parameters (see \S4 below). In order to both efficiently explore the broad parameter space and take small steps to sample the narrow posteriors, we employ a parallel tempering strategy, as discussed, e.g., in ~\citet{Earl2005}. In this approach, the posteriors are sampled at various levels, i.e., replicas, each of which has been artificially broadened by some {\em a priori\/} chosen temperature. Each successive replica in the ladder has an increasing temperature, i.e., corresponding to a larger broadening. Only the lowest level replica, which corresponds to no broadening, is used for the final calculation of the posteriors. However, the other replicas help sample the parameter space more efficiently and increase the chance that a Markov-Chain step will get closer to and sample a region of high posterior. The number and spacing of the parallel tempered ladders (the replicas) we use depends on the complexity of the data and of the model. For the examples with the synthetic data we use in this paper, we typically use a ladder of 40 replicas with temperatures increasing from the lowest replica in multiples of $\sqrt{2}$ from unity.

\begin{figure}[t]
  \centerline{  \includegraphics[width=0.49\textwidth]{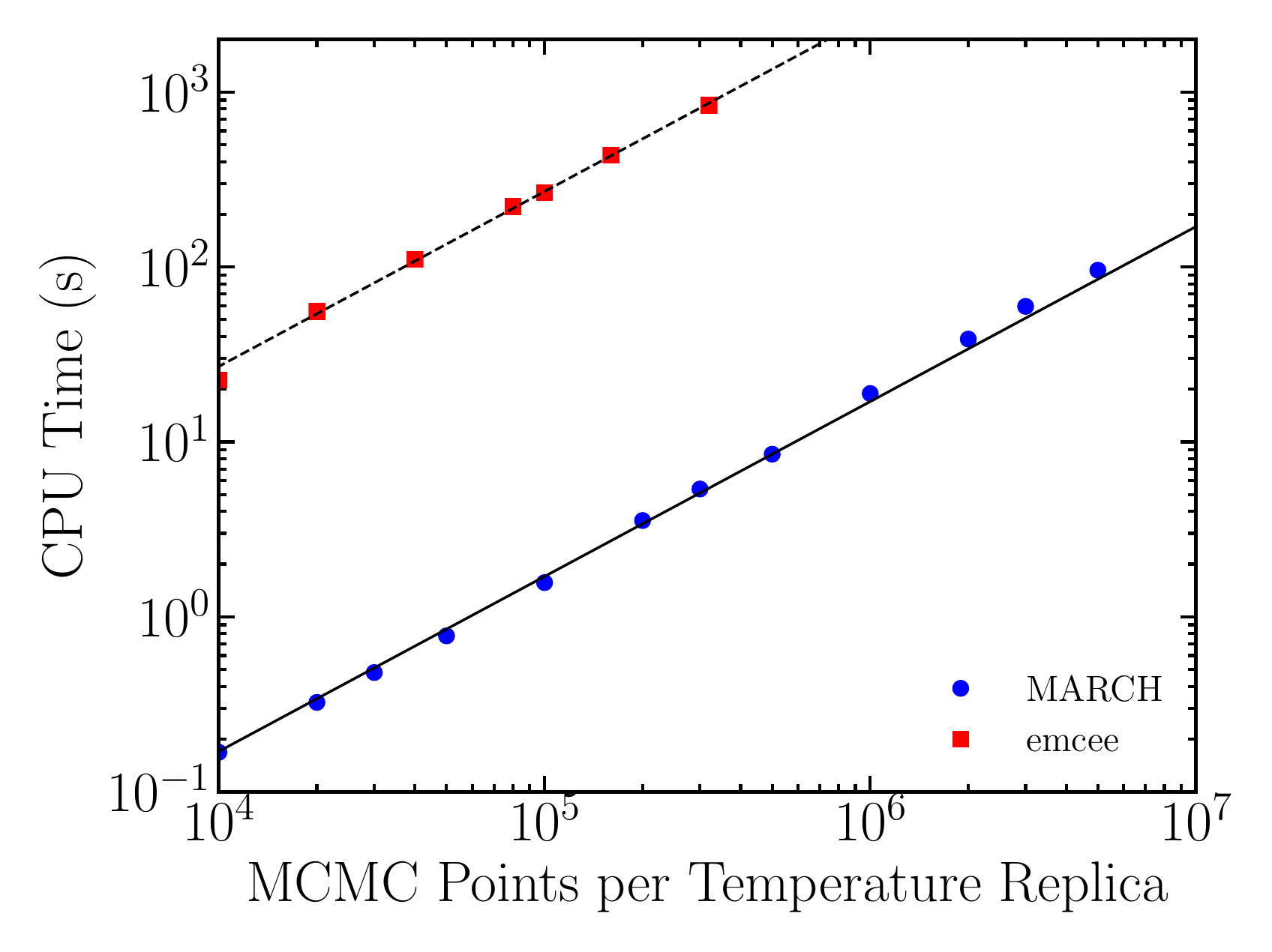}}
    \caption{The CPU time for the execution of the \march\ algorithm on a single CPU, as a function of the number of MCMC chain steps per temperature replica for the synthetic data problem discussed in \S3. The red points show the corresponding performance of the commonly used Python \texttt{emcee} algorithm. \march\ is optimized to achieve high performance even on a single CPU, which reaches $\sim 17$~seconds per $10^6$~MCMC steps for this example.
 \label{fig:perf}}
\end{figure}

Like all MCMC algorithms, \march\ is trivially parallelizable. However, special care was given in its development to boost its computational efficiency and performance in order to enable fitting complex, numerical models to the interferometric data. In particular, we ensured that computationally expensive operations (such as logs, powers, and trigonometric functions) are only sparsely used and organized the algorithmic flow in such a way that the same quantity is not repeatedly computed. As a result, for analytic models, \march\ achieves very high performance even on a single CPU. 

Figure~\ref{fig:perf} shows the CPU time (on a Macbook Air with an 1.6 GHz Intel Core i5 processor) per temperature replica as a function of MCMC chain steps, for the synthetic data problem discussed in \S3 below. In this example, 152 data points were fit with an analytic crescent model described by 5 model parameters, with the MCMC stepping per parameter adjusted such that the average acceptance ratio of the zero-temperature chain is equal to 20\%. The figure demonstrates the linear scaling of the algorithm on the number of MCMC chain steps and its high performance, which reaches (for this example) $\simeq 17$~seconds per $10^6$~MCMC steps. The figure also compares the performance of \march\ to that of a commonly used Python MCMC algorithm (\texttt{emcee};~\citealt{emcee2013}), running the identical problem. The fact that \march\ was optimized for the problem under consideration leads to an increase in performance by a factor of $\sim 150$ compared to the more general \texttt{emcee} algorithm.

\section{An Analytic Model of Crescent Images}

In order to explore some of the characteristics of fitting models to the EHT visibility data, we will employ here the geometric model of a crescent image devised by~\citet{Kamruddin2013}. This model creates a crescent by subtracting the brightness of two uniform but displaced disks. Additional complexities to the model were added in \citet{Benkevitch2016}, who allowed for a gradient in the brightness across the crescent, and in \citet{PaperVI}, where a number of displaced Gaussian components were included to the image. Albeit simple, this model captures all the basic characteristics of the images encoded in the first EHT data on M87~(see~\citealt{PaperVI}). 

The complex visibility of a simple crescent image is given by~\citep{Kamruddin2013}
\begin{equation}
V(u,v)=2\pi I_0 \left[\frac{R_p J_1(2\pi k R_p)}{2\pi k}-e^{-2\pi i (au+bv)}(1-f)\frac{R_n J_1(2\pi k R_n)}{2\pi k}\right]\;,
\label{eq:ring}
\end{equation} 
where $R_p$ and $R_n$ are the outer and inner disk angular radii, $J_1(x)$ is the Bessel function of the first kind, $I_0$ is the uniform brightness of the ring, $f$ measures the fractional depression of  brightness at its center,  
\begin{equation}
k\equiv \sqrt{u^2+v^2}\;,
\end{equation}
is the baseline length in units of the wavelength and the coordinate pair $(a,b)$ measures the displacement of the center of the inner disk with respect to that of the outer disk along the two orientations in the sky.

\begin{figure}[t]
  \centerline{  \includegraphics[width=0.49\textwidth]{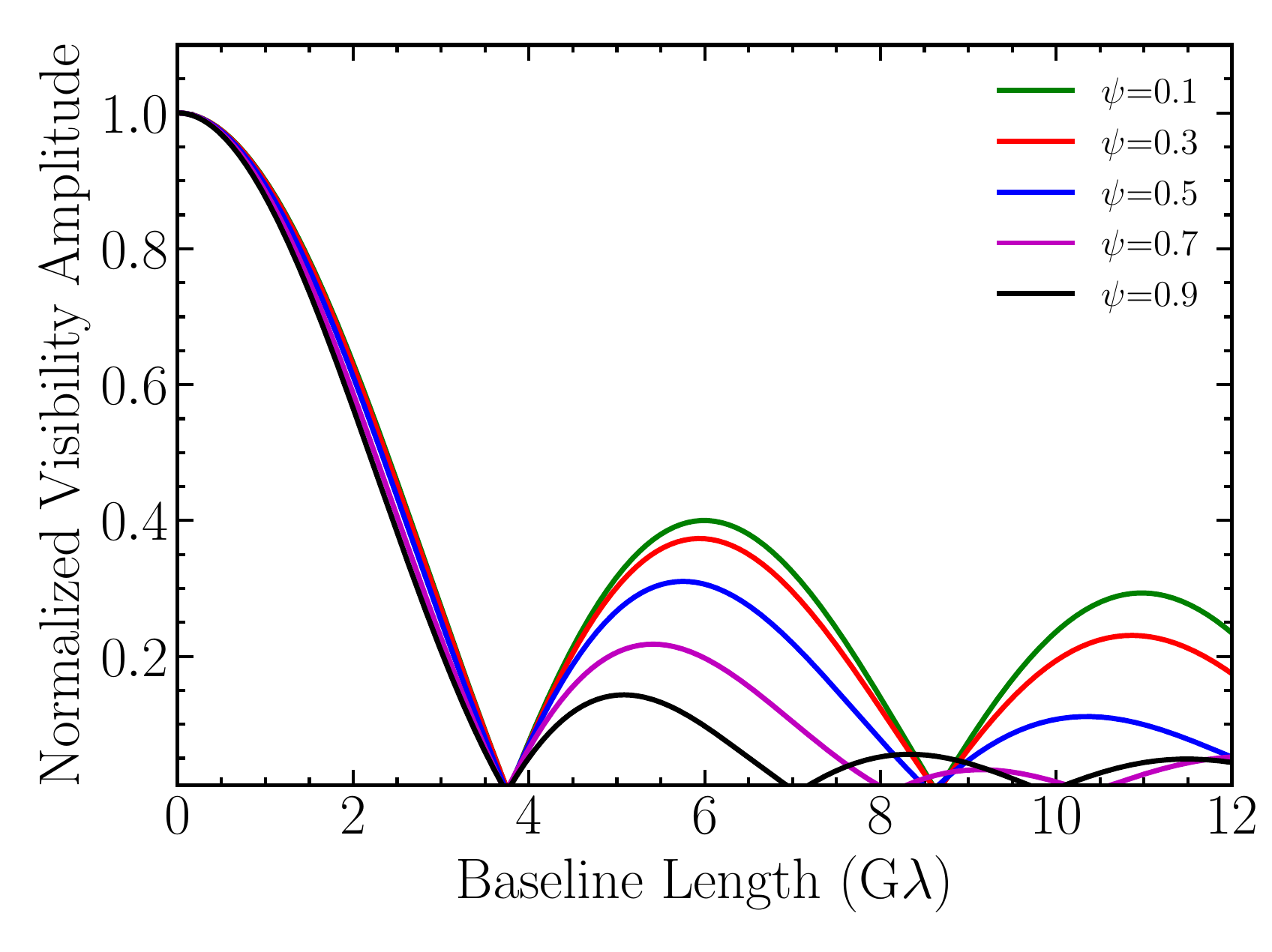}}
    \caption{Normalized visibility amplitude as a function of baseline length for symmetric rings with different fractional widths $\psi$ and radii such that they all show the first minimum at the same baseline length. The radius of the ring with infinitesimal width is set to $R=21~\mu$as, to resemble the 2017 EHT image of M87. The fractional width of the ring determines the relative amplitude of the maximum around $\sim 6$~G$\lambda$ and, for $\psi\gtrsim 0.5$, the location of the second minimum.
 \label{fig:ring_width}}
\end{figure}

In order to simplify our notation, we express the characteristic radius of the crescent in terms of the average radii of the two uniform disks
\begin{equation}
R\equiv \frac{R_p+R_n}{2}
\end{equation}
and the fractional width of the ring as their normalized difference\footnote{Note that, in~\citet{PaperVI}, the fractional width is defined in terms of the mean diameter, i.e., $f_w\equiv (R_p-R_n)/(2R)$, such that $f_w= \psi/(2-\psi)$.}
\begin{equation}
\psi\equiv \frac{R_p-R_n}{R_p}\;.
\end{equation}
For non-symmetric crescents (i.e., when $a\ne 0$ or $b\ne 0$), we  define the orientation of the crescent as
\begin{equation}
\phi\equiv \tan^{-1}\left(\frac{b}{a}\right)
\end{equation}
and the symmetry of the crescent as 
\begin{equation}
\tau\equiv 1-\frac{\sqrt{a^2+b^2}}{R_p - R_n}\;.
\end{equation}
Finally, we normalize the complex visibilities such that they are equal to unity at zero baseline length, i.e., set 
\begin{equation}
I_0=\frac{1}{\pi R_p^2-(1-f)\pi R_n^2}\;.
\end{equation}
Unless otherwise stated, we will set $f=0$ hereafter.

\begin{figure}[t]
  \centerline{  \includegraphics[width=0.49\textwidth]{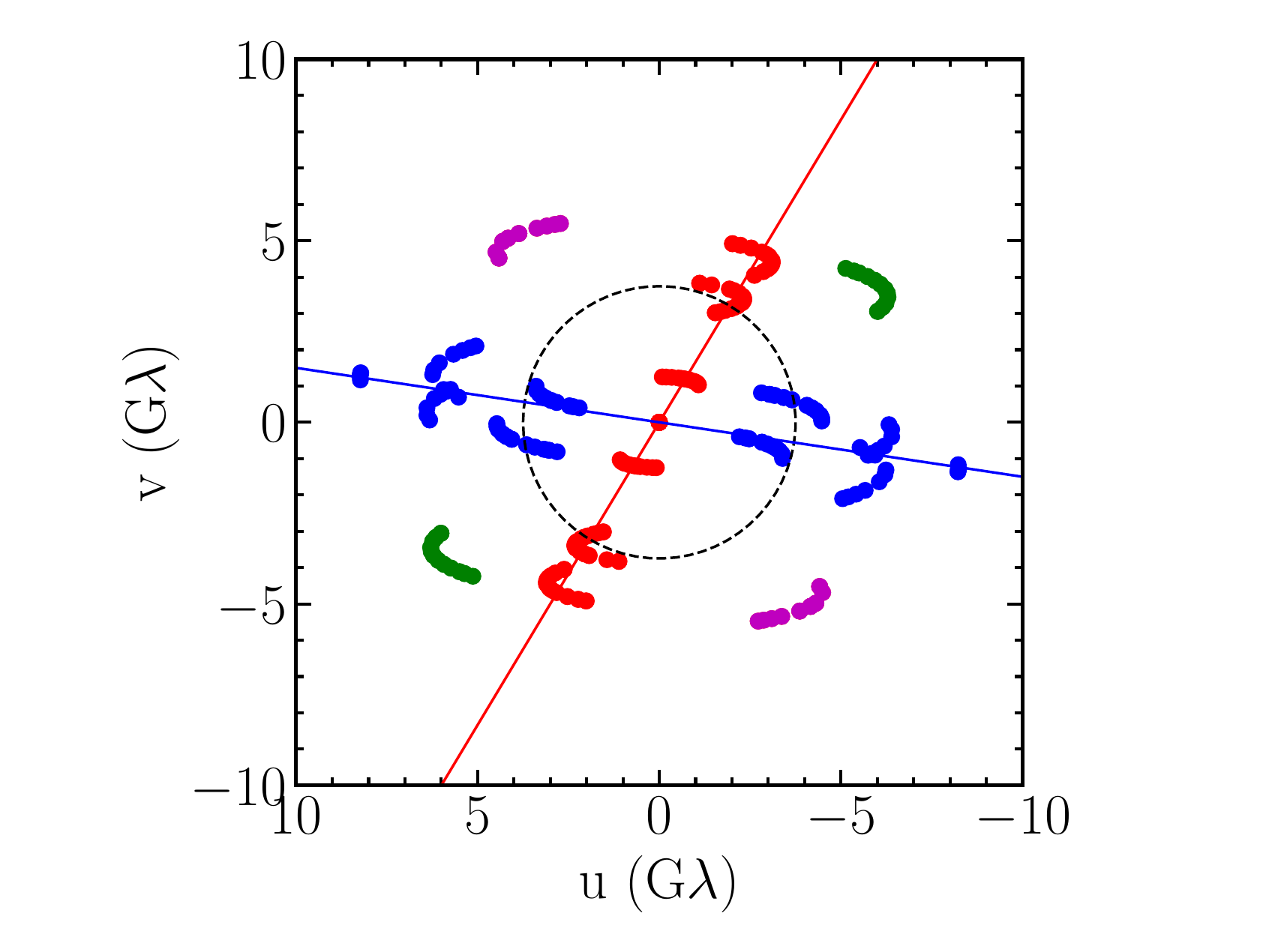}}
    \caption{The $(u,v)$ coverage used in the synthetic data of the symmetric crescent model discussed in \S3. The $(u-v)$ coverage is comparable to that of the 2017 EHT observations of M87. The dashed circle shows the baseline length of the location of the first minimum in visibility amplitude.
    \label{fig:test1_uv}}
\end{figure}

The visibility amplitudes~(\ref{eq:ring}) of symmetric rings show the typical "ringing" of the Bessel functions, with minima at a series of characteristic baseline lengths. For a ring of infinitesimal width, the first minimum occurs at a baseline length of $k_0\simeq 0.383/R$. As the fractional width of the ring increases, the baseline length of the first minimum also increases. This introduces a correlation between the radii and fractional widths of rings that show the first visibility minimum at the same baseline length. An approximate fitting formula for this correlation is~(see~\citealt{PaperVI})
\begin{equation}
R(\psi)=R_0 \frac{1-\psi/2}{1-0.48 \psi +0.11 \psi^5}\;,
\label{eq:Rcor}
\end{equation}
where $R_0$ is the radius of the infinitesimal ring with the same location of the first visibility minimum. Setting $0\le \psi\le 1$, we get $0.64 R_0 \le R(\psi)\le R_0$, for rings that show the first visibility minimum at the same baseline length.

Figure~\ref{fig:ring_width} shows the normalized visibility amplitudes of symmetric rings with different fractional widths but with radii that follow the correlation given by equation~(\ref{eq:Rcor}). By construction, all these visibility amplitudes have the first visibility minimum at the same baseline length ($\simeq 3.75$~G$\lambda$). As the fractional width of the ring increases, the local maximum at $\simeq 6$~G$\lambda$ decreases and, for $\psi\gtrsim 0.5$, the location of the second minimum moves towards smaller baseline lengths. 

\begin{figure}[t]
  \centerline{  \includegraphics[width=0.49\textwidth]{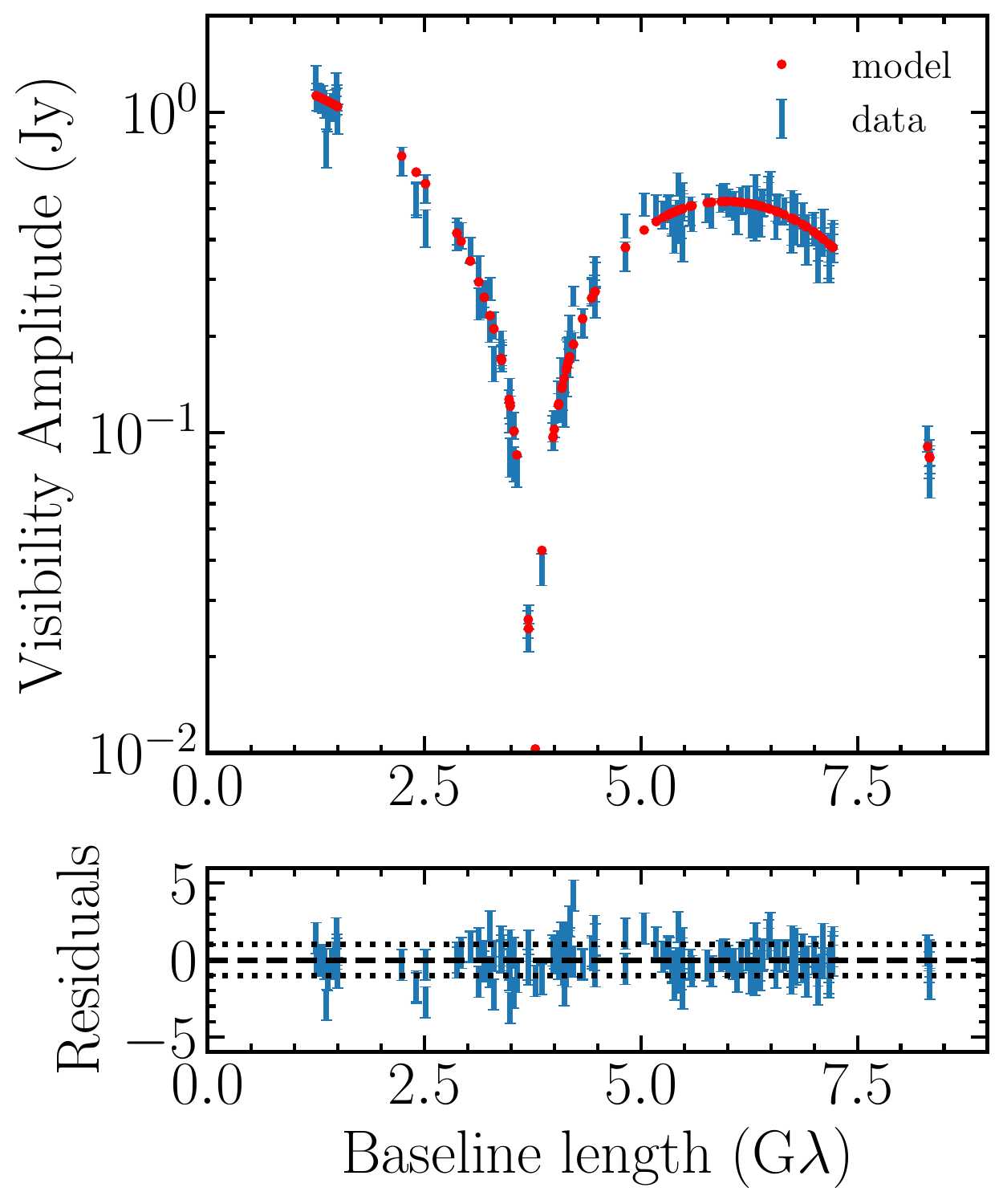}}
    \caption{Synthetic data (blue error bars) and best-fit model (red points) for a symmetric crescent image with $R=21~\mu$as, $\psi=0.1$, and the $(u-v)$ coverage of the 2017 EHT observations of M87 shown in Figure~\ref{fig:test1_uv}. For this test, the uncertainties are purely thermal (Gaussian) with a standard deviation equal to 10\% of the model visibility amplitudes. The residuals are calculated as [(model-data)/error].
 \label{fig:test1_visamp}}
\end{figure}

In order to explore the performance of our MCMC algorithm and understand the qualitative aspects of fitting crescent models to EHT data, we generated a set of synthetic visibility amplitude and closure phase data using the crescent model~(\ref{eq:ring}) for a symmetric ring ($a=b=0$) with parameters $R=21~\mu$as, $\psi=0.1$, $\phi=0$, $\tau=1$, and a flux of $1.3$~Jy. We generated data on the $(u,v)$ locations of the baselines of the 2017 April 5 EHT observations of M87 (see Fig.~\ref{fig:test1_uv}) and, for the purposes of this exploration, introduced only thermal (Gaussian) errors with standard deviation equal to 10\% of the model visibility amplitudes. We then used \texttt{MARCH} to fit the data with the same crescent model. The synthetic visibility amplitudes, the predictions of the best-fit model, and their residuals are shown in Figure~\ref{fig:test1_visamp}.

\begin{figure}[t]
  \centerline{  \includegraphics[width=0.49\textwidth]{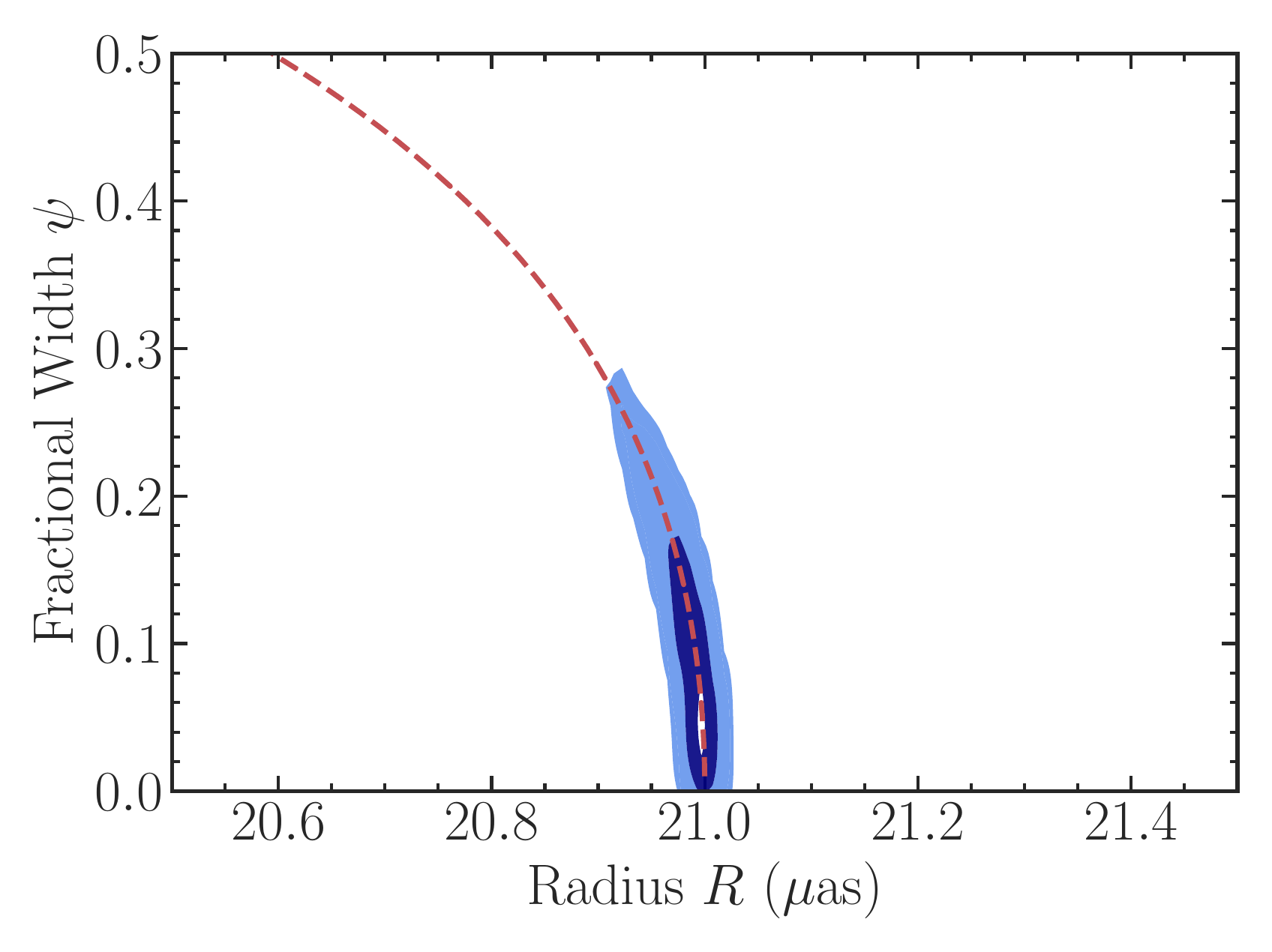}}
    \caption{The correlated 68\% and 95\% credible intervals in the inferred radius $R$ and fractional width $\psi$ for the synthetic data shown in Figure~\ref{fig:test1_visamp}. The dashed line shows the expected correlation between the two model parameters such that the visibility amplitudes show a minimum at a baseline length of $\simeq 3.75$~G$\lambda$. It is the baseline length of this first minimum that almost single-handedly determines the most likely model parameters and their correlations.
        \label{fig:test1_corr}}
\end{figure}

Figure~\ref{fig:test1_corr} shows the 68\% and 95\% credible intervals for the two primary parameters of the model: the radius $R$ of the ring and its fractional width $\psi$. The values of these parameters and their correlation can be understood completely in terms of the location of the minimum in the visibility amplitudes that occurs at a baseline length of $\simeq 3.75$~G$\lambda$ (see Fig.~\ref{fig:test1_visamp}). The red dashed line in Figure~\ref{fig:test1_corr} shows the expected correlation between fractional width and radius for all symmetric rings that have the first minimum of the visibility amplitude occurring at that baseline length. (Note that, for the purposes of this figure, which requires a relative accuracy of better than 1 part in 1000, we used the numerical solution for the correlation between fractional width $\psi$ and ring radius $R$ and not the analytic approximation~[\ref{eq:Rcor}]).

Marginalizing the posterior over all model parameters other than the radius results in $R=20.98^{+0.02}_{-0.03}$, which corresponds to a fractional uncertainty of 0.1\%. This is a surprisingly small uncertainty given that there are 152 visibility amplitude and 119 closure phase data points, the uncertainty of each complex visibility was set to 10\%, and the model has only a handful of free parameters. For comparison, if this was a power-law model in visibility, i.e., $V(b)=A b^c$, and there were $N=152+119=271$ equidistant data points in $\log(b)$ between a minimum baseline $b_{\rm min}=1$~G$\lambda$ and a maximum baseline $b_{\rm max}=8$~G$\lambda$, all with fractional uncertainty $\sigma_V/V$, then the uncertainty in the coefficient $c$ of that model would be 
\begin{eqnarray}
\sigma_{c}&\simeq &\left(\frac{12}{N}\right)^{1/2} \frac{\sigma}{\ln(b_{\rm max}/b_{\rm min})}\nonumber\\
&\simeq & 1.0\left(\frac{N}{271}\right)^{-1/2} \left(\frac{\sigma_V/V}{10\%}\right) \%\;.
\end{eqnarray}
i.e., an order of magnitude larger. 

The fact that the uncertainties in the measurement of one of the key model parameters is substantially smaller than what one might expect suggests that the model predictions for at least some of the data points depend very strongly on that model parameter. This might be viewed naively as good news, since it allows for strong constraints to be placed on model parameters with only marginal uncertainties in the data. However, it also implies that the inferred uncertainties in the model parameters rely extremely strongly on the ability of the model to predict accurately the salient features of the visibility data. The latter is unlikely to be true when we are using simple geometric models (e.g., crescent shapes) to fit visibility data that describe images of turbulent accretion flows around black holes. 

\section{The Role of Influential Data Points}

The example discussed in the previous section demonstrates that a very small number of data points near the location of salient features in the visibility data obtained with an interferometer such as the EHT have an uneven degree of influence on the measurements and uncertainties of model parameters. This is true even for our example that was based on synthetic data generated by the actual crescent model, with the same fractional uncertainties for each data point. The effect gets exacerbated when the data points have different uncertainties (i.e., they are heteroscedastic) and the actual data are only broadly described by the simple geometric model that we try to fit to them, as is the case in real applications to EHT data. 

Quantifying the degree of influence of different data points on model fitting has been extensively explored in both the frequentist (see, e.g.,~\citealt{Cook1977,Cook1979}) and the Bayesian frameworks~(see, e.g., \citealt{Millar2007}). Both approaches generate very similar outcomes as they identify two types of influential data points. The first type are the outliers, i.e., the ones that deviate from model predictions by a degree that is statistically inconsistent with the uncertainties. The presence of such data points alters the inferred uncertainties in the model parameters. The second type are the subset of data points that depend substantially more strongly on the model parameters than the remaining data points and, therefore, exert uneven leverage in determining the best-fit values of the model parameters.

\begin{figure}[t]
  \centerline{  \includegraphics[width=0.49\textwidth]{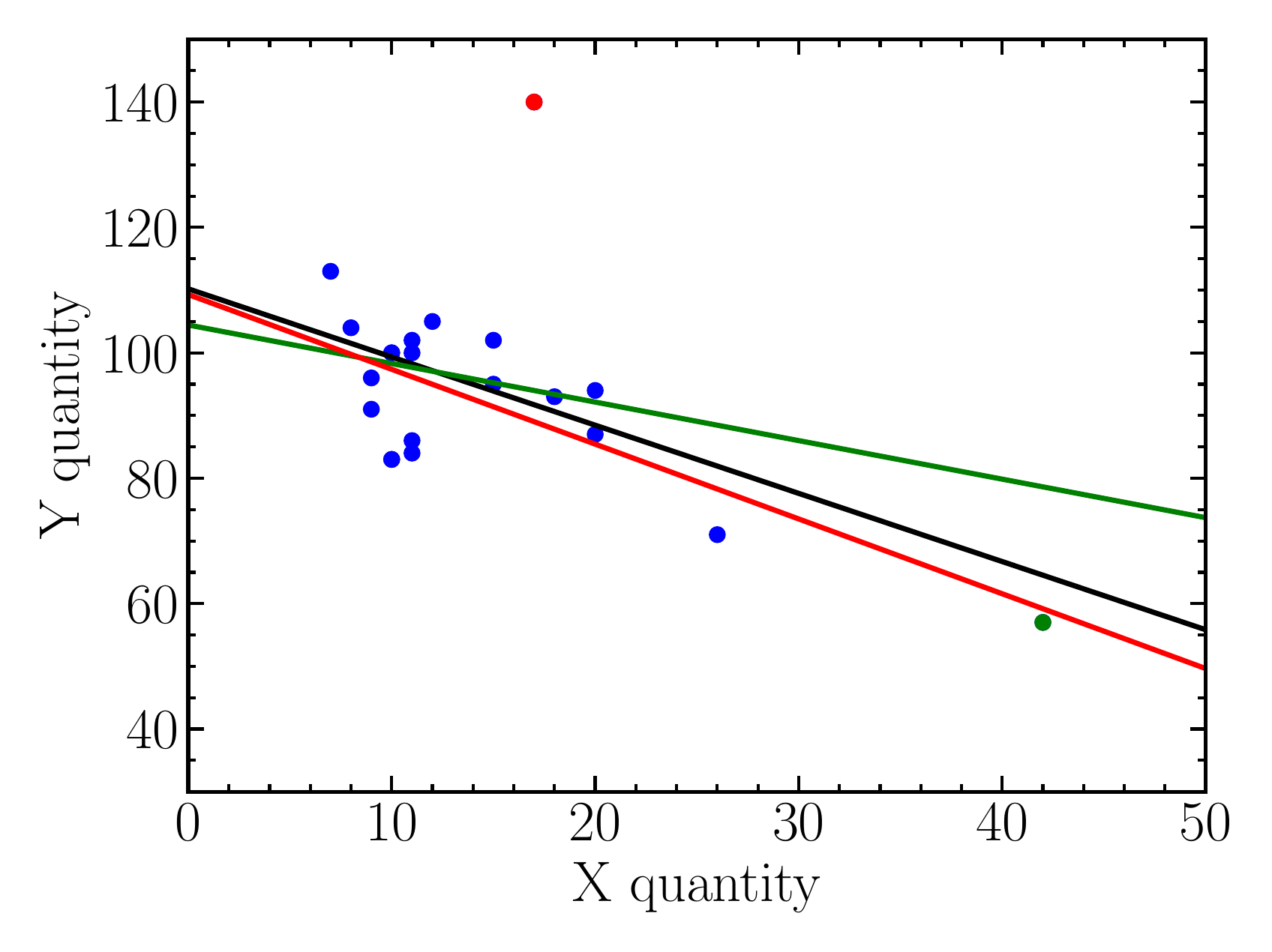}}
    \caption{A toy example of the effect of influential points on linear regression (based on the data set discussed in~\citealt{Millar2007}). The black line is the best-fit linear model to all the data points. The red and green lines are the best-fit models after removing the red or the green point, respectively. The model parameters are given in Table~\ref{table:infl}. The red data point is an outlier and its presence increases substantially the inferred uncertainties of the model parameters. The green data point, because of its location, introduces a significantly larger leverage in determining the model parameters than the remaining data points. 
        \label{fig:fit_infl}}
\end{figure}

A toy example with both types of data points is shown in Figure~\ref{fig:fit_infl}, where a linear model is fit to the data set used in the Bayesian analysis of~\citet{Millar2007}; the best-fit line is shown in black and its parameters are shown in Table~\ref{table:infl}. The red data point is a clear outlier. Removing it introduces only a small change to the parameters of the best-fit line (red curve); however, removing just that single point reduces the inferred uncertainties of the model parameters by as much as a factor of 2. The green data point appears at a large horizontal distance from the other data points and, therefore, exerts substantial leverage in determining the model parameters. Removing the green point changes substantially the best-fit model parameters, i.e., leads to a best-fit line (green curve) that is significantly flatter than when the full data set is used. This toy example demonstrates how removing even a single influential point changes substantially the parameters of the best-fit model and their uncertainties. For this reason, identifying such influential points for EHT data is critical in assessing the robustness of the results of model fits.

\begin{deluxetable}{lcc}
\tablewidth{800pt}
\tablecolumns{3}
\tablecaption{Linear Model Parameters (Fig.~\ref{fig:fit_infl})\label{table:infl}}
\tablehead{Data Set & slope & y-intercept}
\startdata
Full Data Set &  -1.09 $\pm$ 0.40  &  110.2 $\pm$ 6.5\\
Red Point Excluded & -1.19 $\pm$ 0.24 & 109.3 $\pm$ 4.0\\
Green Point Excluded & -0.61 $\pm$ 0.66 & 104.4 $\pm$ 9.2\\
\enddata
\end{deluxetable}

\subsection{Identifying Influential Data Points}

In order to quantify and formally measure the influence of each data point on results of model fiting, we follow the Bayesian approach of~\citet{Millar2007}. The qualitative idea is to measure the change in the information content of the resulting posterior when we artificially broaden (or shrink) the likelihood that corresponds to each one of the data points. If this change is similar among all data points, this would imply that all of them have similar influences on the result. If, on the other hand, this change is substantially larger for some of the data points, we will conclude that the latter have high influence on the result.

We define the quasi-likelihood ${\cal L}($data$\vert \vec{\theta};w_i)$ as the likelihood of obtaining the data set from a particular set of model parameters $\vec{\theta}$ but with the likelihood of the $i-$th data point having been broadened by a factor $w_i$, i.e., (cf. eq.~[\ref{eq:like}])
\begin{equation}
{\cal L}({\rm data}\vert \vec{\theta};w_i) \equiv {\cal L}_i({\rm data}\vert\vec{\theta})^{w_i}
\prod_{j\ne i}  {\cal L}_j({\rm data}\vert\vec{\theta})\;.
\label{eq:like_wi}
\end{equation}
If the likelihood of the $i-$th data point is a Gaussian with dispersion $\sigma_i$, the contribution of this data point to the total likelihood in equation~(\ref{eq:like_wi}) is also a Gaussian but with a dispersion equal to $\sigma_i/w_i^{1/2}$. When $w_i\rightarrow 0$, the $i$-th data point does not contribute to the determination of the best-fit parameters.

Using this quasi-likelihood, we define a new quasi-posterior by effectively elevating the parameter $w_i$ to a model parameter as (cf.\ eq.~[\ref{eq:bayes}])
\begin{equation}
P(\vec{\theta}\vert{\rm data};w_i)=C\; P_{\rm pri}(\vec{\theta})\; {\cal L}({\rm data}|\vec{\theta};w_i) \;;
\label{eq:bayes_wi}
\end{equation}
here we have incorporated the prior $P_{\rm pri}(\vec{\theta})$ on the model parameters. When $w_i=1$, the above quasi-posterior is equal to the true posterior~(\ref{eq:bayes}). In order to quantify the change in the information content of the posterior when we artificially change the likelihood of the $i$-th data point, i.e., when we set $w_i=1\pm \epsilon$, we use the Kullback-Leibler (KL) divergence
\begin{eqnarray}
&&D_{KL,i}\left[P(\vec{\theta}\vert{\rm data};w_i), P(\vec{\theta}\vert{\rm data};1)\right]
\nonumber\\
&&\qquad\qquad=
\int d\vec{\theta}\;P(\vec{\theta}\vert{\rm data};w_i) \log\left[ \frac{P(\vec{\theta}\vert{\rm data};w_i)}{P(\vec{\theta}\vert{\rm data};1)}\right]\;.
\end{eqnarray}
By definition, when $w_i=1$, the KL divergence is equal to zero as is its first derivative with respect to $w_i$, under normal conditions~\citep{Millar2007}. The second derivative of the KL divergence with respect to $w_i$ offers the lowest-order non trivial measure of the variation with $w_i$. After some simple manipulations, the second derivative becomes
\begin{equation}
\ddot{D}_{KL,i}(\vec{\theta}\vert {\rm data})=\int d\vec{\theta}\; P(\vec{\theta}\vert {\rm data})
\left[\left.\frac{\partial \log P(\vec{\theta}\vert{\rm data};w_i)}{\partial w_i}\right\vert_{w_i=1}\right]^2\;.
\label{eq:DKL}
\end{equation}
This is the well known result that the curvature of the KL-divergence is equal to the Fisher information with respect to the relevant parameter (here $w_i$).

Using equations~(\ref{eq:like_wi}), (\ref{eq:bayes_wi}), and (\ref{eq:DKL}) we obtain
\begin{equation}
\ddot{D}_{KL,i}(\vec{\theta}\vert {\rm data})=\int d\vec{\theta}\; P(\vec{\theta}\vert {\rm data})
\left[\log {\cal L}_i(\vec{\theta}\vert{\rm data})\right]^2\;.
\end{equation}
This is the measure of the influence of the $i$-th data point on the results and it is equal to the variance of the log-likelihood of this data point across the space of the model parameters. \cite{Millar2007} suggest as a normalized measure of the influence of the $i$-th data point the ratio
\begin{equation}
M_i\equiv \frac{\int d\vec{\theta}\; P(\vec{\theta}\vert {\rm data})
\left[\log {\cal L}_i(\vec{\theta}\vert{\rm data})\right]^2}{\int d\vec{\theta}\; P(\vec{\theta}\vert {\rm data})
\left[\log {\cal L}(\vec{\theta}\vert{\rm data})\right]^2}\;,
\label{eq:influence}
\end{equation}
where the denominator involves the combined log-likelihood for all data points. In general, the sum $\sum_i M_i$ will not be equal to unity, because the individual likelihoods may be correlated. Also note that \citet{Spiegelhalter2002} (see also \citealt{Gelman2013}) identified the denominator of equation~(\ref{eq:influence}) as a measure of the complexity of the model given the data, i.e., of the effective number of degrees of freedom. Among a set of $N$ data points, those with $M_i\gg 1/N$ will be the ones with disproportionate influence on the result of the model fitting. 

\begin{figure}[t]
  \centerline{  \includegraphics[width=0.49\textwidth]{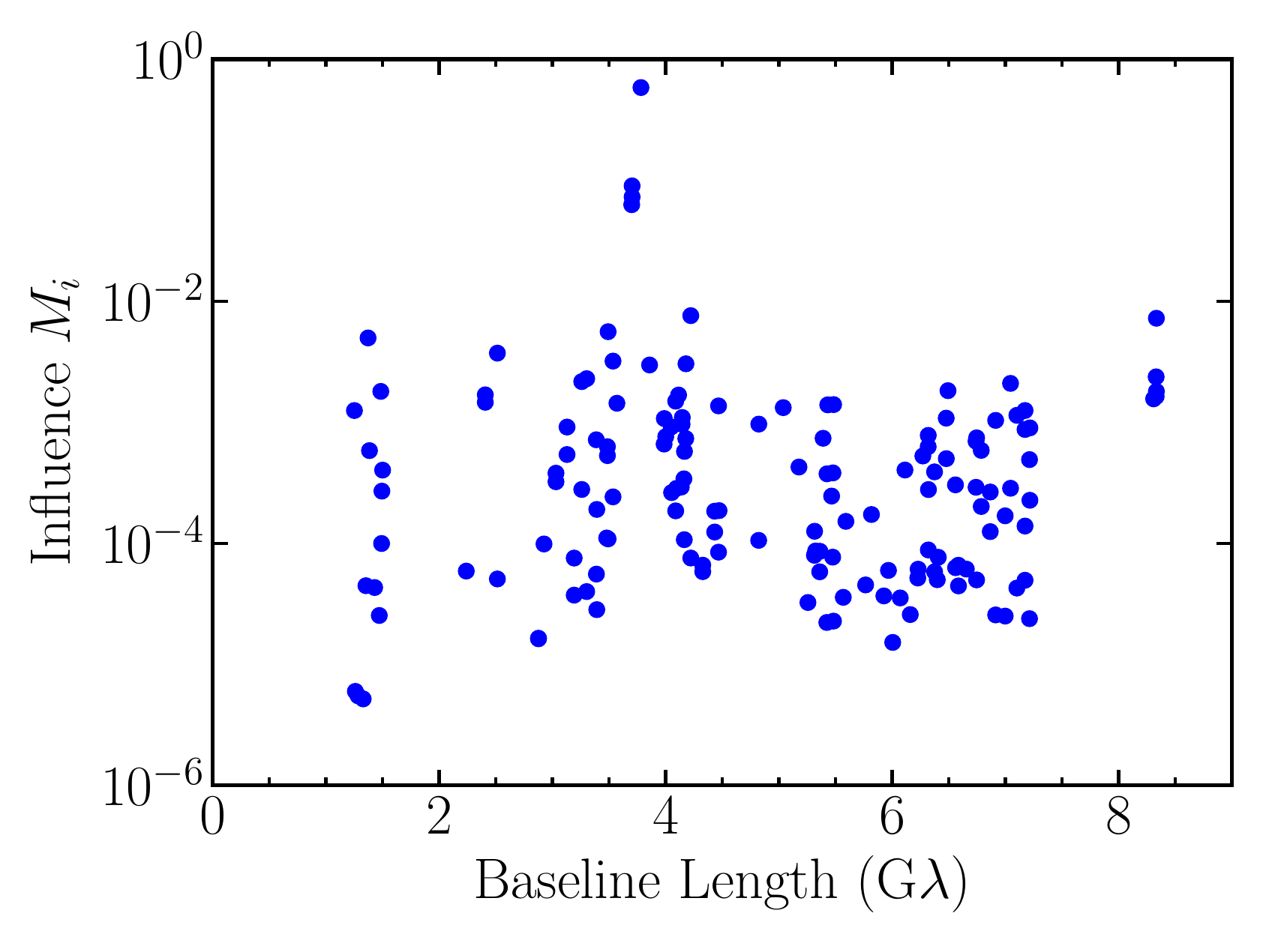}}
    \caption{The measure of the influence $M_i$ of each data point on the outcome of fitting a crescent model to the synthetic EHT data shown in Figure~\ref{fig:test1_visamp}, plotted as a function of the baseline length that corresponds to each point. The three data points near the location of the minimum in visibility amplitude at $\simeq 3.5$~G$\lambda$ exert disproportionate influence on the result of the model fitting. 
    \label{fig:test1_infl}}
\end{figure}

\subsection{Influential Data Points in Crescent Images}

In this subsection, we apply the formalism we developed in \S4.1 to the test problem of fitting crescent models to interferometric data that we discussed in \S3 (see Figures~\ref{fig:test1_visamp}-\ref{fig:test1_corr}). Figure~\ref{fig:test1_infl} shows the influence $M_i$ (eq.~[\ref{eq:influence}]) of each of the $N=152$ visibility-amplitude data points on the result as a function of its baseline length. For the vast majority of data points, $M_i\lesssim 1/N = 1/152$. This is not true, however, for the three data points that lie at the first minimum of the visibility amplitude curve. The large values of $M_i$ for these three data points, which are several orders of magnitude larger than the others, demonstrate their disproportionate influence on the result. Indeed, the fact discussed in \S3 that the best-fit values of and correlations between the inferred radius and fractional width of the crescent are determined almost entirely by the location of the first minimum in the visibility amplitude data is directly related to this disproportionate influence. Note that the influence values ${\cal M}_i$ of the corresponding closure-phase data do not single out any additional influential points for this synthetic data set. 

\begin{figure}[t]
  \centerline{  \includegraphics[width=0.49\textwidth]{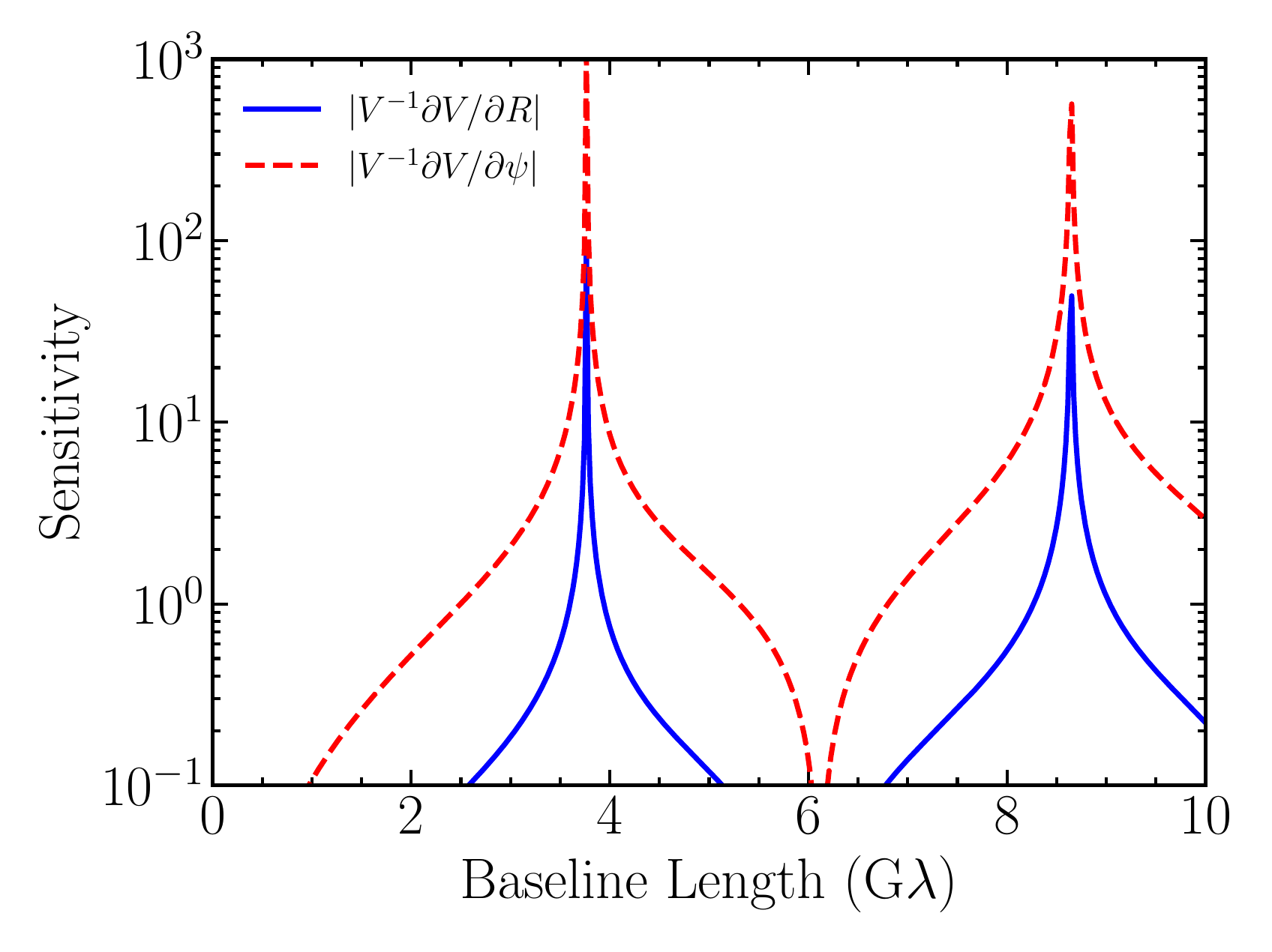}}
    \caption{The sensitivity of predicted visibility amplitudes $V$ for a symmetric ring model on two of the model parameters, the radius $R$ (solid blue line) and the fractional width $\psi$ (dashed red line) of the ring, as a function of baseline length. For this example, the radius is set to $R=21~\mu$as and the fractional width to $\psi=0.1$. The peak sensitivity occurs, as expected, at baseline lengths that correspond to minima in the visibility amplitude.
    \label{fig:sensitivity}}
\end{figure}

In order to understand the origin of the disproportionate influence of the data points near the visibility minima on the outcome of the model comparison, we explore the sensitivity of the inference of model parameters on the location of data points of the $u-v$ plane. As we did above, we focus on the visibility amplitude data, although similar considerations apply to the closure phase data as well. The sensitivity in the inference of a particular model parameter $\theta_i$  on the measurement at the location ($u_j,v_j$) of a given data point will depend on the uncertainty $\sigma_j$ of the measurement (the higher the uncertainty the weaker the sensitivity) and on the derivative of the model prediction with respect to $\theta_i$, evaluated at the location of the data point, i.e.,
\begin{equation}
\left.\sigma_j^{-1} \frac{\partial V}{\partial \theta_i}\right\vert_{u_j,v_j} = \left(\frac{\sigma_j}{V_j}\right)^{-1}
\left.V_j^{-1} \frac{\partial V}{\partial \theta_i}\right\vert_{u_j,v_j}\;.
\end{equation}
This term can be recognized as the contribution of the $j$-th data point on the Cram\'er-Rao bound of the uncertainty in the inference of the $i$-model parameter. In the last expression, we have normalized the measurement uncertainty of each data point by the model predictions $V_j$ in order to be consistent with our synthetic data that have constant fractional measurement uncertainties. With this normalization, a sensitivity value of unity implies that making, e.g., a change of 1\% to the model parameter will cause a change of 1\% to the model prediction.

\begin{figure}[t]
  \centerline{  \includegraphics[width=0.49\textwidth]{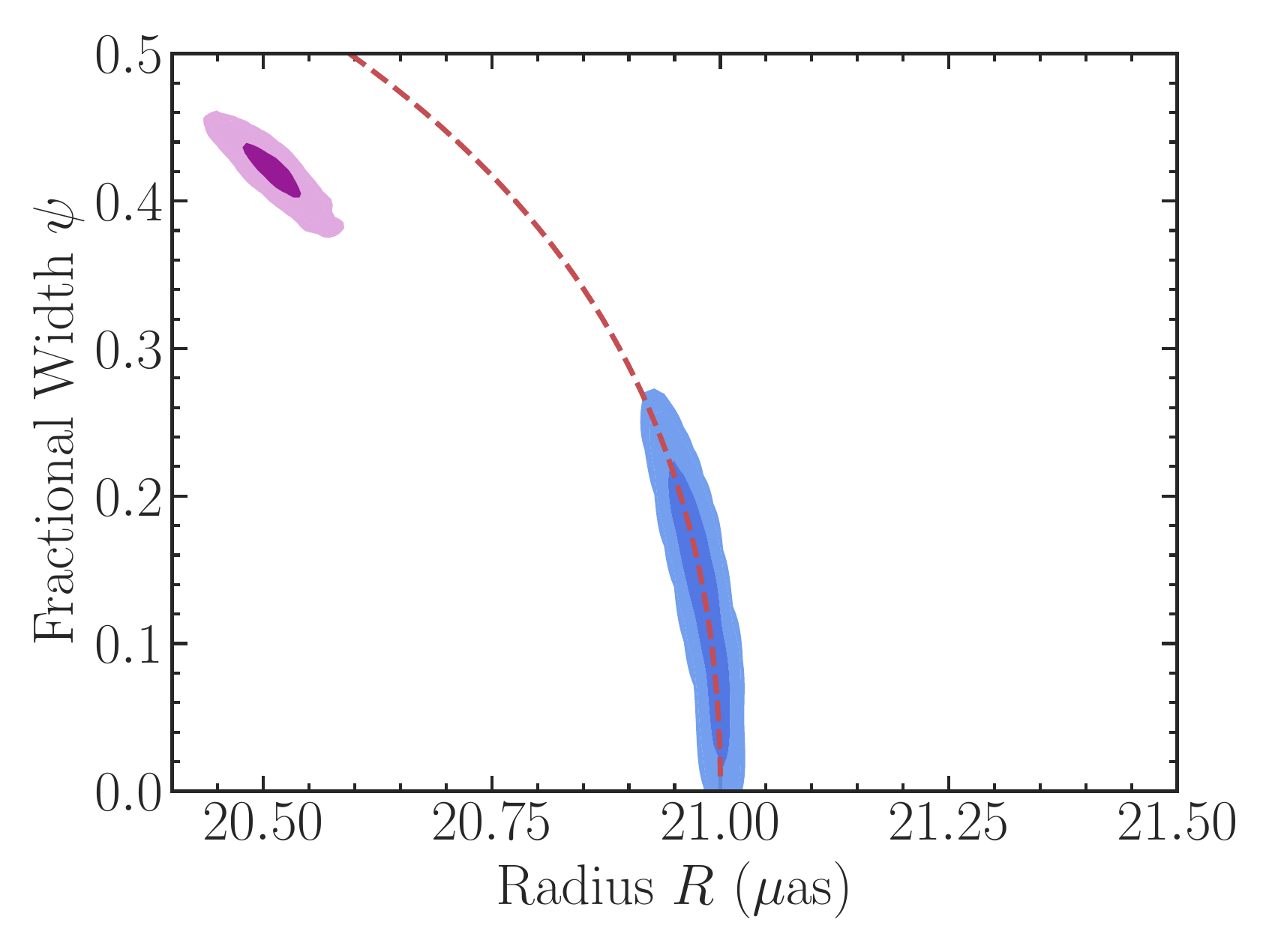}}
    \caption{As in Figure~\ref{fig:test1_corr} but including, in magenta, the posteriors for the synthetic data in which a point source was added to the center of the ring image carrying a flux of 0.02~Jy; the parameters of the ring remain the same. Even though this point source carries $\sim 1.5$\% of the total flux of the image and it affects predominantly the complex visibilities of the handful of data points near the visibility minima, it modifies the inferred model parameters by a degree that is substantially larger than the statistical uncertainties. This is a direct consequence of the disproportionate influence of the data points near the visibility minima on the fit.
        \label{fig:test3_corr}}
\end{figure}

Figure~\ref{fig:sensitivity} shows the sensitivity of inferring the radius and fractional width of the symmetric ring model using data points with constant fractional uncertainties, as a function of the baseline length of each data point. As expected, both sensitivity curves show very strong peaks near the locations of the minima in the visibility amplitude. Moreover, the inference of the fractional width of the ring has a higher sensitivity on the measurements than the inference of its radius. The location and magnitude of the peaks in these two sensitivity curves match the locations and relative magnitudes of the influence $M_i$ of the various data points shown in Figure~\ref{fig:test1_infl}. 

To illustrate the potential biases introduced by the disproportionate influence of a small number of data points on the model parameters, we generate a new set of synthetic data from the ring model with the same parameters and $u-v$ coverage as those discussed in \S3 and shown in Figures~\ref{fig:test1_uv}-\ref{fig:test1_corr}. We add to this ring model, however, a single point source at the center of the image that carries a flux of only 0.02~Jy. This is about 1.5\% of the total flux of the image and effectively adds a constant $\simeq 0.02~$ to all visibility amplitudes, thus disproportionately affecting the complex visibilities of the data points near the minima in the visibility amplitude curve. The change introduced to the complex visibilities of the vast majority of the remaining data points is within their statistical uncertainties. We then attempt to fit this new synthetic data set with a crescent model that does not include this weak point source. The goal of this test is to simulate the outcome of trying to fit a simple model to an image that has marginally more complex structures than the model allows.

\begin{figure}[t]
  \centerline{  \includegraphics[width=0.49\textwidth]{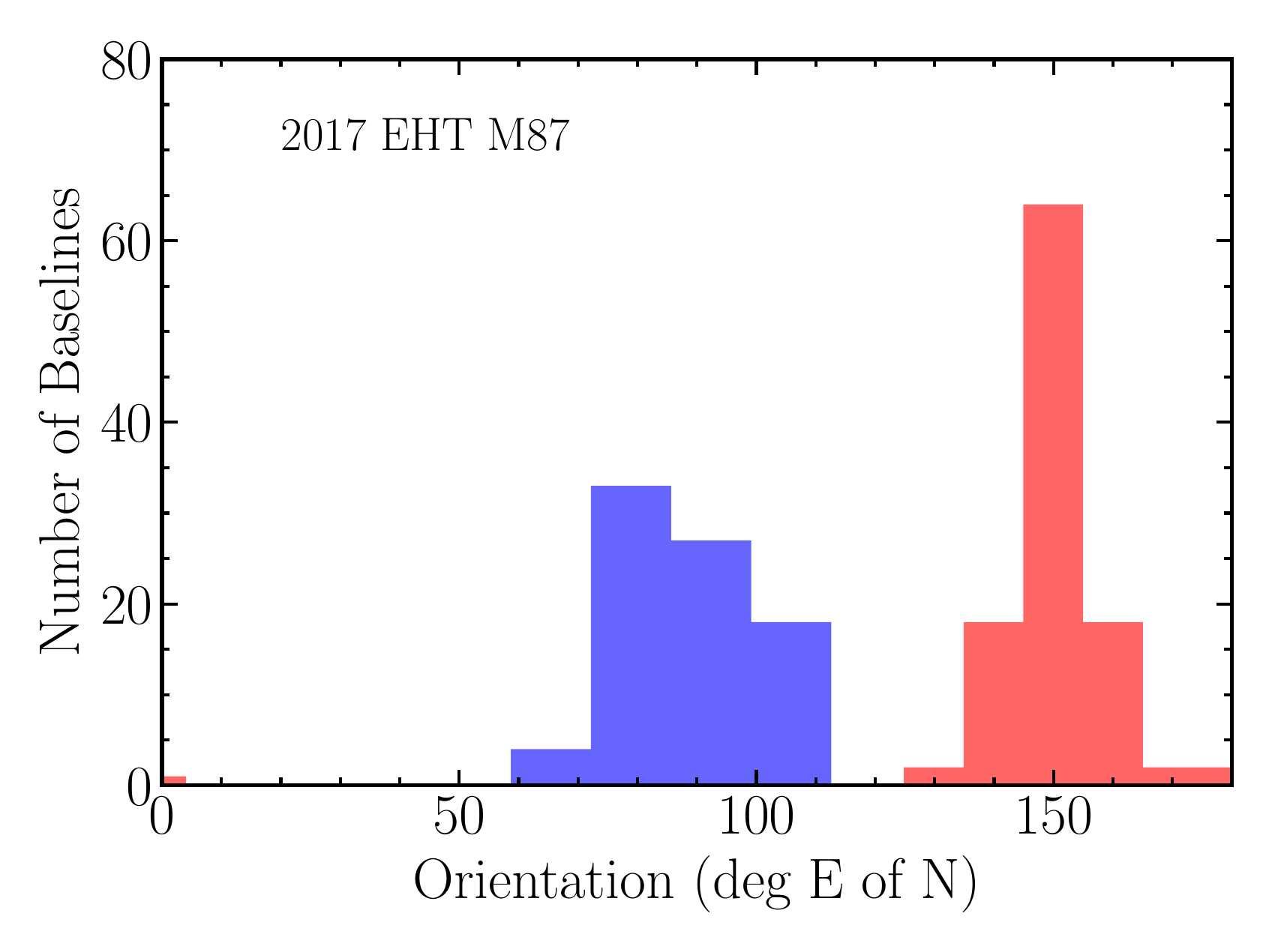}}
    \caption{The distribution of orientation angles (measured in degrees East of North) of the red- and blue-color coded EHT baselines in the 2017 observations of M87 as shown in Fig.~\ref{fig:test1_uv}. 
   As can be also seen from the $u-v$ map in the latter figure, most baselines fall primarily along E-W orientations with position angles $\sim 80-100^\circ$ and along N-S with position angles $\sim 150^\circ$.
        \label{fig:eht_coverage}}
\end{figure}

The result for the radius and fractional width for this revised data set are shown in Figure~\ref{fig:test3_corr}. Even though the addition of the weak point source affects only a handful of interferometric data points, the high degree of influence of these data points causes a bias to the inferred model parameters to a degree that is substantially larger than the formal uncertainties from the fitting process. The correlated credible levels for the two parameters do not include the ring parameters that were assumed in the synthetic data. They fall near but not on the expected correlation shown in the red dashed line, given the location of the minimum in the visibility amplitude. As expected from our discussion of the sensitivity of the various model parameters on the measurements, the bias introduced to the fractional width is particularly large (0.4 instead of 0.1 of the underlying image) and is substantially larger than the bias in the radius of the ring.

Note that, in the synthetic dataset we used in these examples, we have set the uncertainties of all data points equal to the same fractional value (10\%). This is not the case for the real EHT data, which are widely heteroscedastic; the signal-to-noise ratio for some ALMA baselines can be as large as 300, whereas for some other baselines involving smaller telescopes it is equal to only a few~(see~\citealt{PaperIII}). In this case, the combination of the leverage of each data point (because of its location) and its signal-to-noise ratio will determine its influence: data points with high signal-to-noise ratios in locations of high leverage will have the largest influence. We will explore this more realistic case in \S6.

\begin{figure}[t]
  \centerline{  \includegraphics[width=0.49\textwidth]{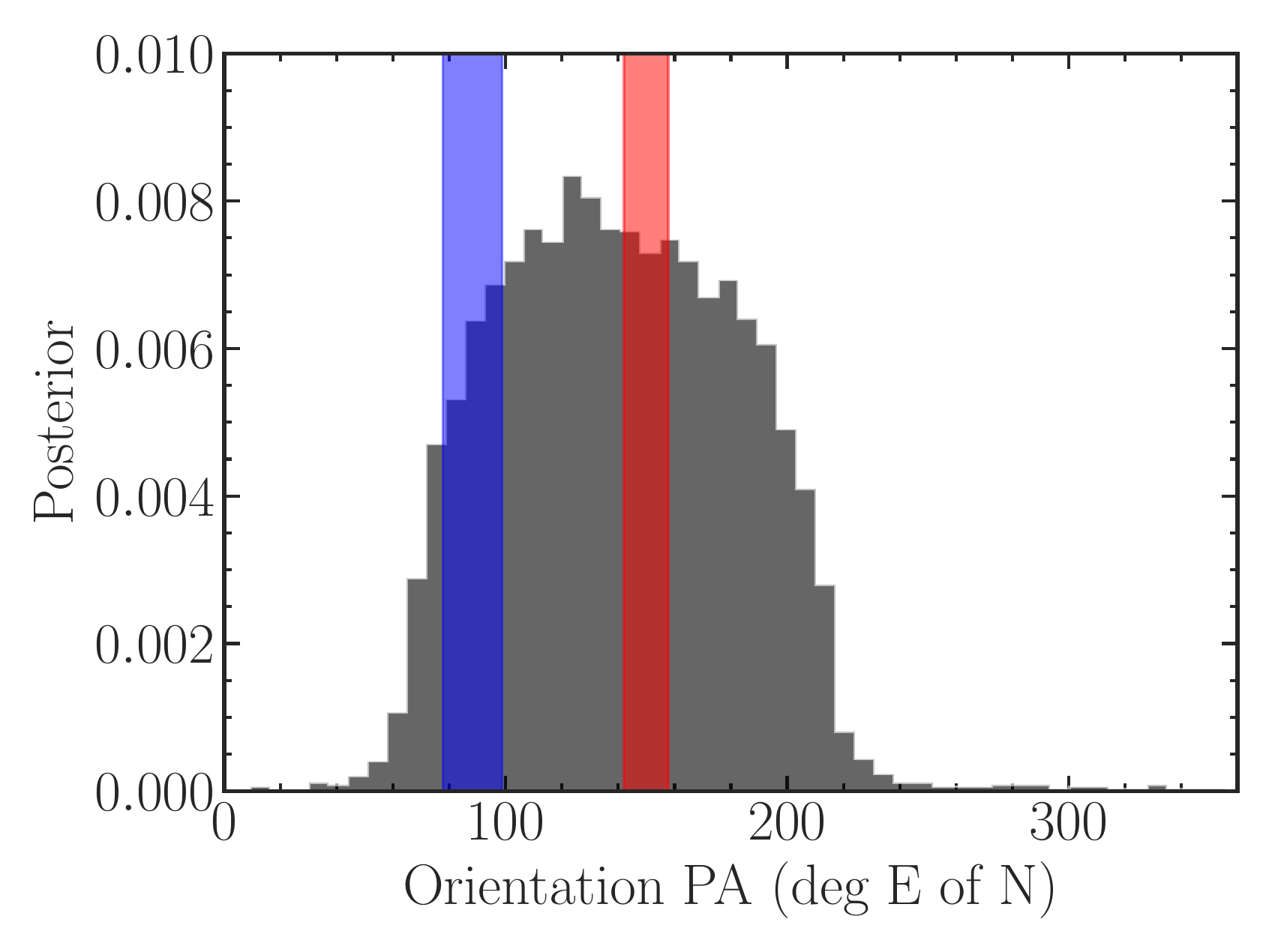}}
    \caption{The posterior over the orientation of a crescent model after fitting to the synthetic data shown in Figure~\ref{fig:test1_visamp}. The vertical blue and red hatched regions show the main orientations of the baselines used in generating the data displayed with the same colors in Fig.~\ref{fig:test1_uv}. Even though the underlying model is a fully symmetric ring, the posterior prefers a slightly asymmetric crescent shape with an axis of symmetry that lines up with the major orientations of baseline coverage in the $u-v$ plane, biasing substantially the measurement.
        \label{fig:test1_angle}}
\end{figure}

\section{The Effect of Sparse Coverage}

A second important feature of the EHT data that affects the outcome of fitting theoretical models to them is the sparse coverage of the interferometric $u-v$ plane (see, e.g., Fig~\ref{fig:test1_uv}). As Figure~\ref{fig:eht_coverage} shows, in the current form of the array, most of the EHT baselines lie primarily along N-S and E-W orientations, with substantial gaps between them, at all baseline lengths. This preferential orientation of the baselines can introduce strong biases in some of the inferred parameters of models that are fit to the EHT data, as we will now show.

\begin{figure*}[t]
  \centerline{ 
   \includegraphics[width=0.33\textwidth]{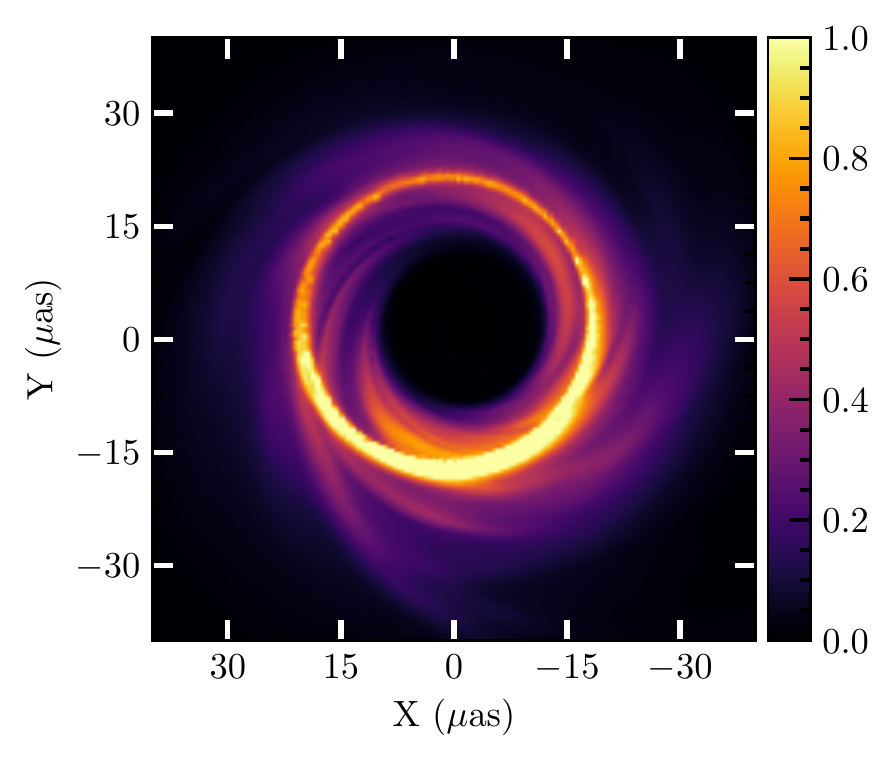}
   \includegraphics[width=0.33\textwidth]{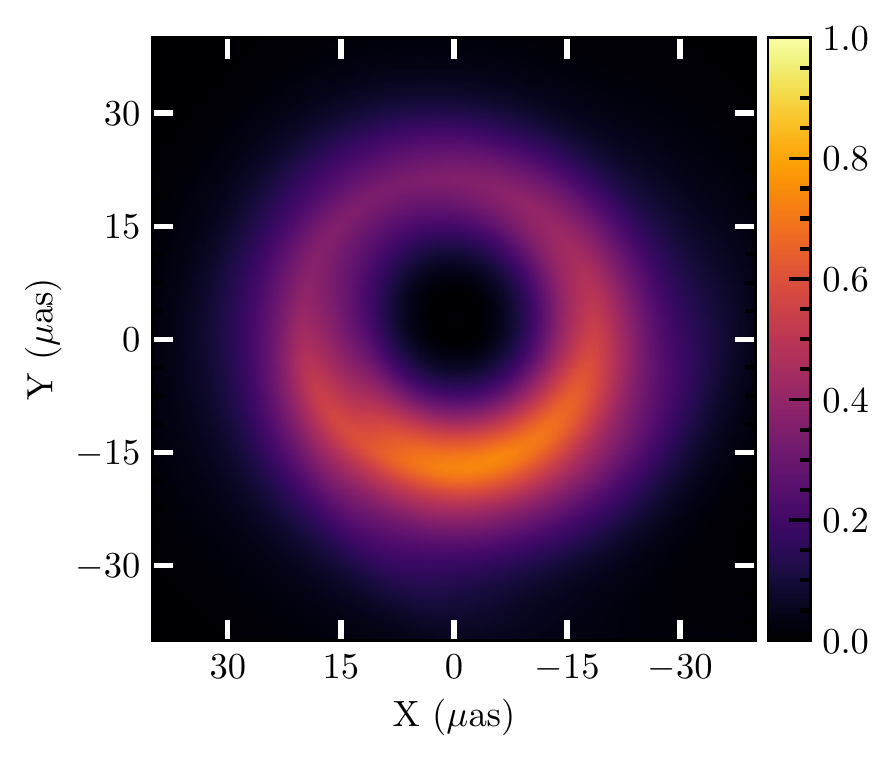}
\includegraphics[width=0.33\textwidth]{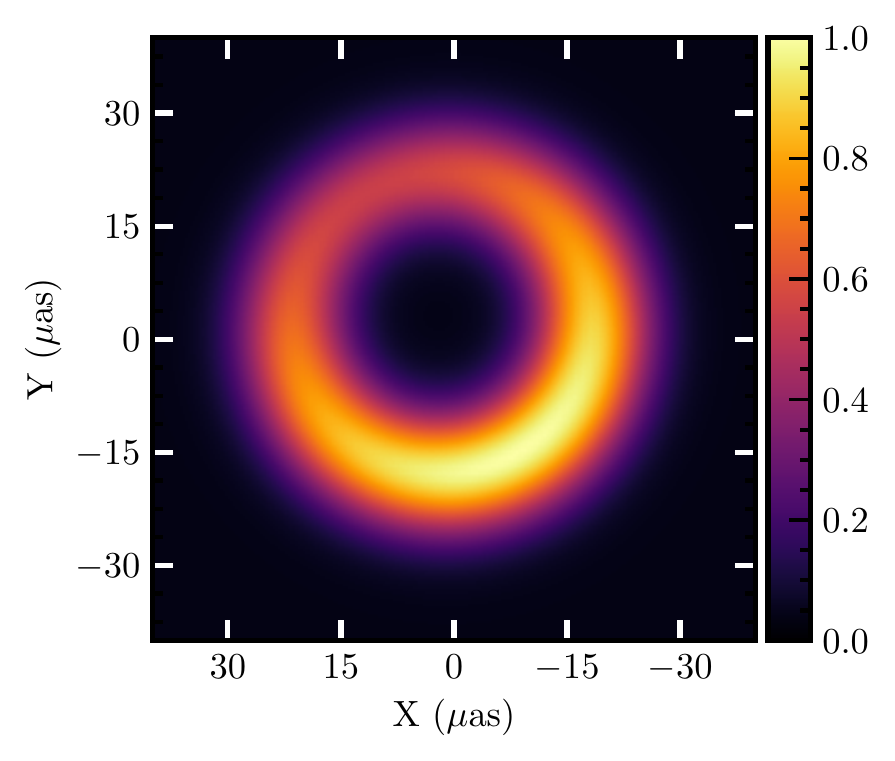}}
    \caption{{\em (Left)\/} Snapshot image from a MAD GRMHD+GR Radiative Transfer simulation with parameters that are relevant to the M87 black hole; this snapshot is dominated by a crescent structure. {\em (Middle)\/} The same snapshot filtered with an $n=2$ Butterworth filter with a cutoff at 15~G$\lambda$. {\em (Right)\/} The best-fit crescent image to the synthetic EHT data generated from the snapshot on the left panel. For this snapshot that is dominated by a crescent structure, fitting a crescent model to synthetic data results in a model image that closely matches the one used to generate the synthetic data.
    }
        \label{fig:MAD_134_image} 
\end{figure*}

\begin{figure*}[t]
  \centerline{  \includegraphics[width=0.4\textwidth]{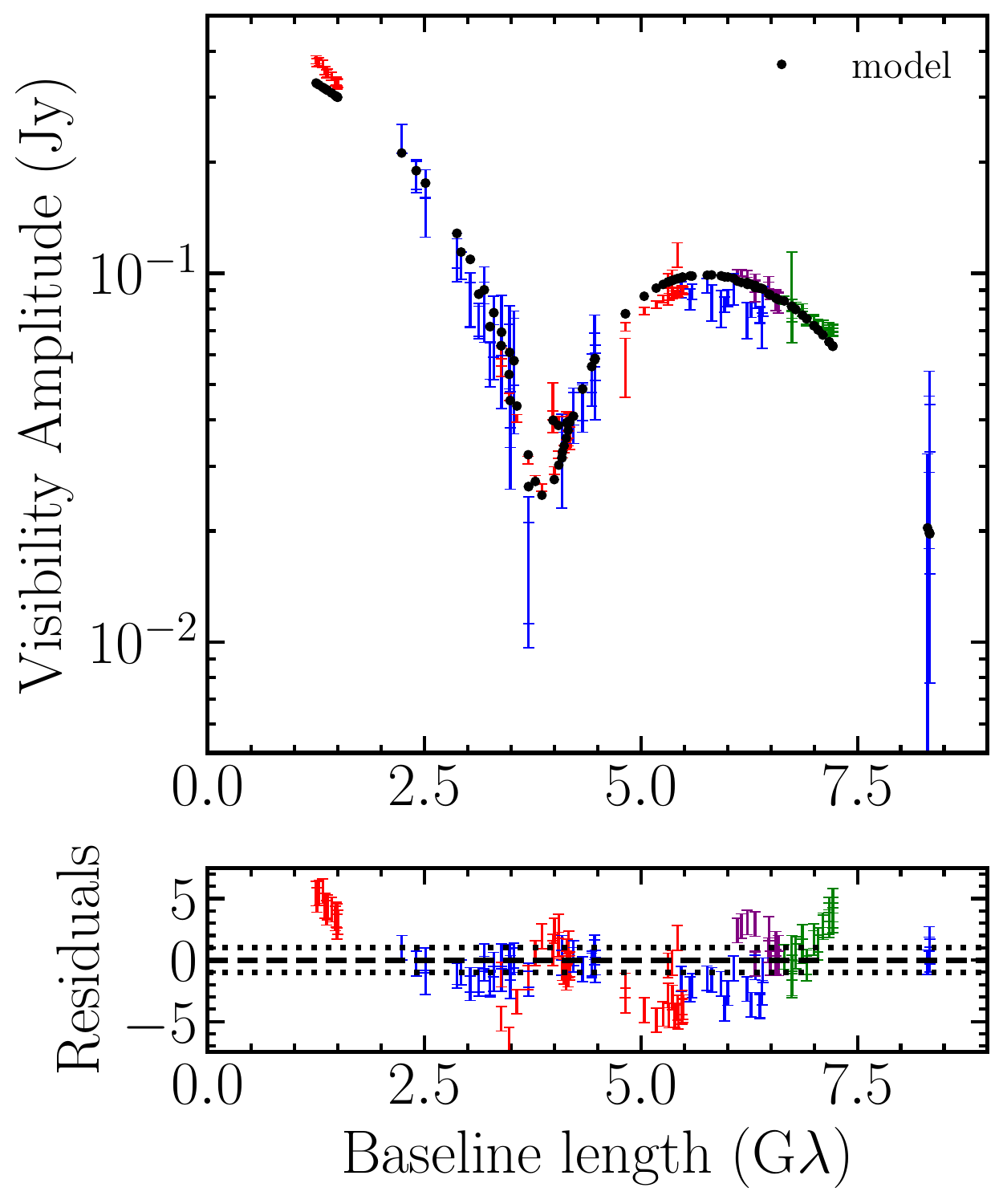}\hspace*{2.0cm}
   \includegraphics[width=0.4\textwidth]{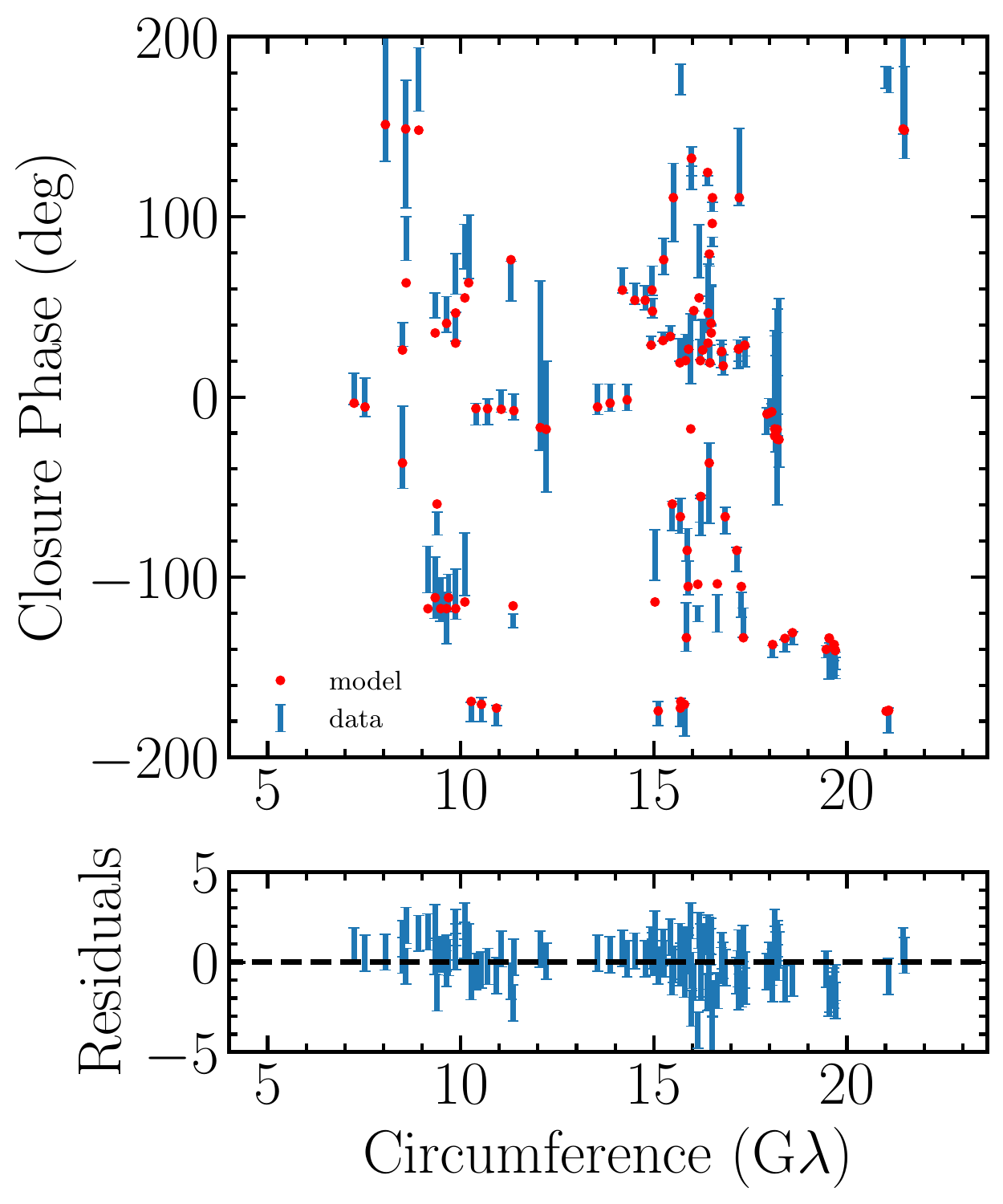}  }
    \caption{Synthetic data (shown with error bars, color coded in the left panel as in Fig.~\ref{fig:test1_uv} according to the orientations of the various baselines) and best-fit model (shown with black points) for fitting a crescent model to synthetic data generated from the image shown in Fig.~\ref{fig:MAD_134_image}. The residuals are calculated as [(model-data)/error]. The left panel shows the visibility amplitude as a function of baseline length and the right panel shows the closure phase as a function of the circumference of the triangle of baselines along which is closure phase was measured. For this snapshot that is dominated by a crescent structure, fitting a crescent model to synthetic data results in only mild residuals.
     \label{fig:MAD_134_visamp}}
\end{figure*}

Figure~\ref{fig:test1_angle} shows the marginalized posterior over position angle of the crescent model fit to the synthetic data introduced in \S3 (see also Fig.~\ref{fig:test1_uv}-\ref{fig:test1_corr}); for consistency with earlier literature, we report the position angle in degrees East of North, i.e., as $90^\circ-\phi$. Even though the underlying image in this example is a fully symmetric ring with no preferred orientation, the posterior over orientation angle peaks at values similar to those of the primary orientations of the EHT baselines. This is a direct consequence of the projection-slice theorem, which states that a cross section of a 2-dimensional Fourier transform along a particular orientation is equal to the Fourier transform of the projection of the image along the same orientation. As discussed in \S3, the analytic crescent model has a preferred orientation along which the cross section of the Fourier transform is independent of the degree of symmetry ($\tau$) of the crescent. In other words, crescents that are mildly different than the symmetric ring (i.e., with a mild asymmetry) will lead to the same predictions as the symmetric model for the observations along the EHT baselines as long as their symmetry axes are lined up with the orientations of the baselines. When marginalized over all other model parameters, the large volume in the space of models occupied by asymmetric crescents that are aligned with the EHT baselines and can reproduce the data within their uncertainties gives rise to a posterior over orientation $\phi$ that is aligned with the EHT baselines.

It is important to emphasize here that not only crescent models but also quasi-analytic (see, e.g., \citealt{Broderick2009}) and even GRMHD model images~(see, e.g., \citealt{PaperV}) of black-hole shadows have intrinsic symmetry axes dictated by the spin of the black hole and the brightness asymmetry introduced by Doppler effects. As a result, similar considerations apply when fitting such models to the EHT data. Moreover, the synthetic dataset we used here is characterized by the same fractional uncertainties for all data points, which is not realistic. In a more realistic situation, a complex image of a black hole will be observed with the EHT and will result in a heteroscedastic dataset. In this case, it will be a complex combination of the preferred orientations of the baselines with the largest signal-to-noise ratio, of the underlying symmetry and orientation of the image, as well as of the symmetry properties of the model that will determine potential biases in the posterior  over orientation angle.

\begin{figure*}[t]
  \centerline{  \includegraphics[width=0.90\textwidth]{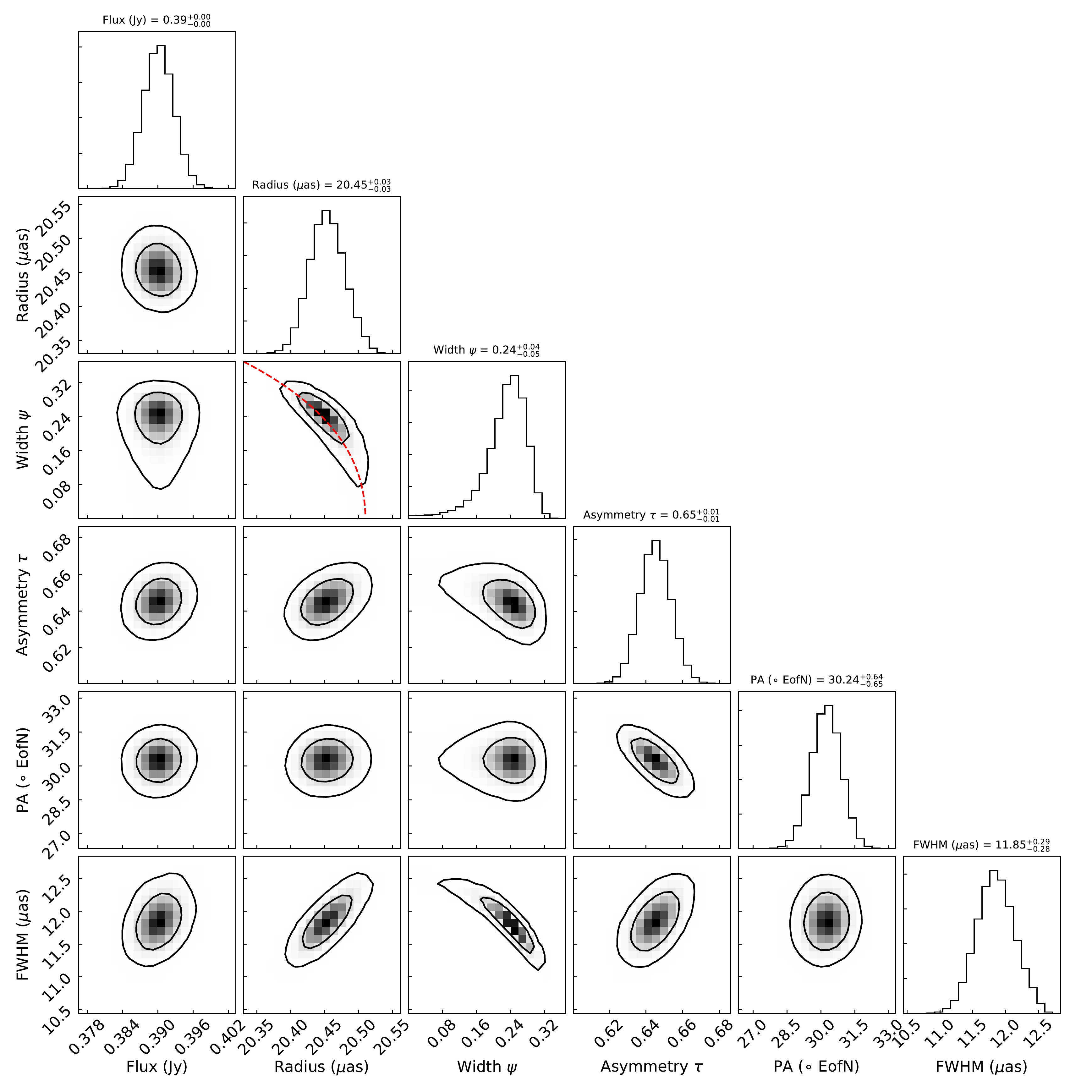}}
    \caption{Posteriors of the various crescent model parameters used to fit the synthetic data from the image shown in Figure~\ref{fig:MAD_134_image}. The red dashed line in the panel showing the correlated posterior between the fractional width $\psi$ and the radius of the crescent corresponds to equation~(\ref{eq:Rcor}) that connects these two parameters for all crescents with the first minimum in the visibility amplitude occurring at the same baseline length.
 \label{fig:MAD_134_corner}}
\end{figure*}

\begin{figure}[t]
  \centerline{  \includegraphics[width=0.45\textwidth]{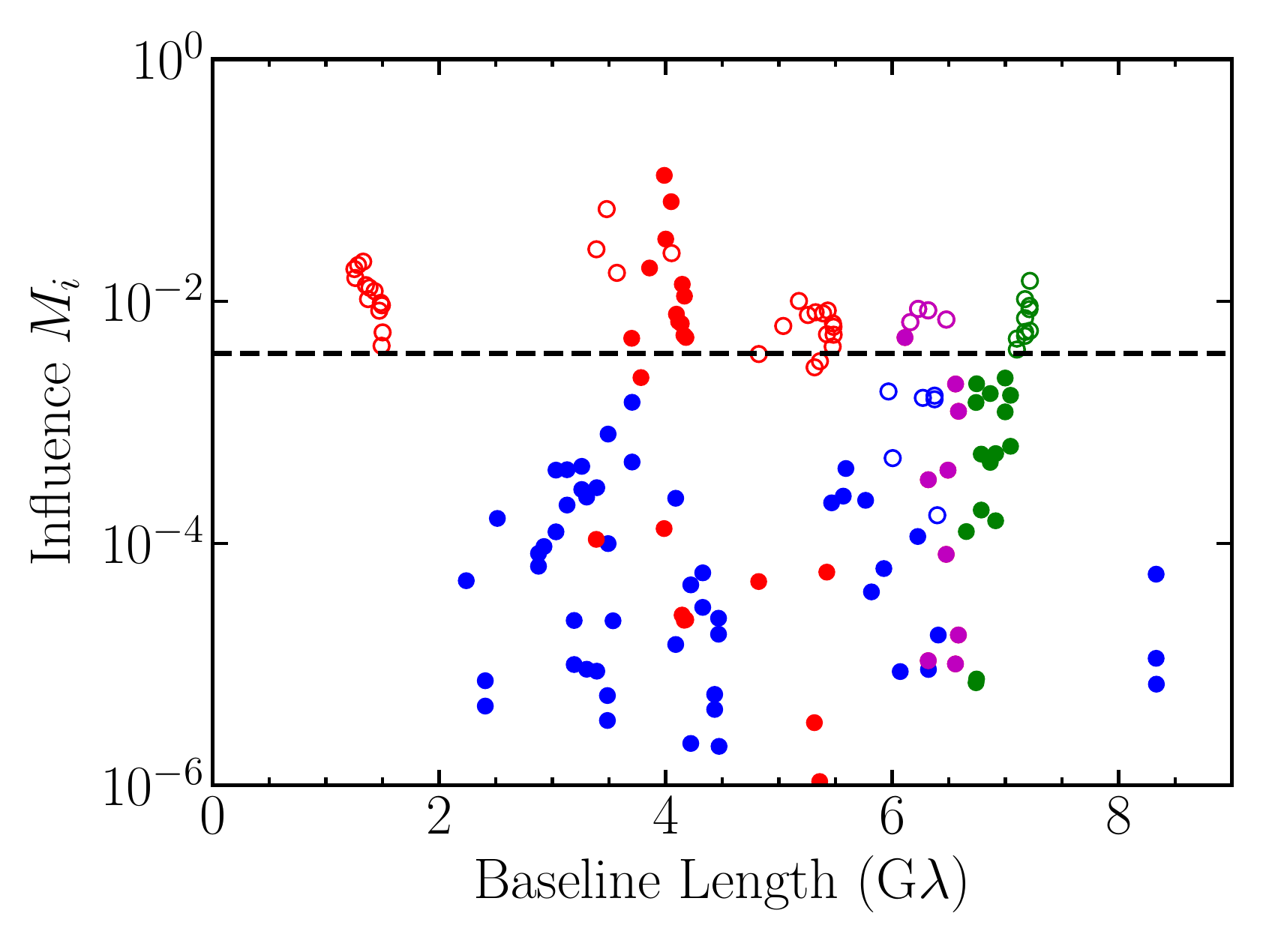}}
    \caption{The influence $M_i$ for each of the synthetic data from the image shown in Figure~\ref{fig:MAD_134_image}. The data points are color-coded as in Fig.~\ref{fig:test1_uv} to show the orientations of the various baselines. Filled circles denote points that are fit well by the crescent model whereas open circles denote points that are not (i.e., the difference between the measurement and the model prediction is less than or more than 2.5 times the measurement uncertainty, respectively). The horizontal dashed line corresponds to $1/N$, where $N$ is the total number of data points and separates the most influential (above the line) from the least influential (below the line) points. The influential data points at $\sim 4$~G$\lambda$ are fit well by the model and have high leverage because they are near the location of the minimum in the visibility amplitude. The other data points above the dashed line are influential because they are outliers.  
     \label{fig:MAD_134_influence}}
\end{figure}

\section{Application to Synthetic EHT Data}

In order to explore a more realistic situation that includes small-scale structure, as well as realistic measurement uncertainties, we used two of the GRMHD+GR Radiative Transfer simulations calculated by~\citet{Chan2015}, but for parameters relevant to the M87 black hole. In particular, we used one simulation with the Magnetically Arrested Disk (MAD)  field structure~\citep{Narayan2012}. We set the mass of the black hole to $M = 6.5\times10^9~M_\odot$, the spin of the black hole to $a=0.9$, the inclination of the observer to $i =17^{\circ}$, the position angle of the black-hole spin axis to 290$^\circ$ East of North, and the electron temperature to follow a plasma $\beta$-dependent prescription with $ R_{\mathrm{high}} = 20$ (see~\citealt{PaperV}). We created images at 1.3~mm using the GPU-accelerated GRay algorithm~\citep{Chan2013} and calculated angular sizes in the sky for a distance of 16.8~Mpc. Finally, we generated synthetic EHT data from images generated with these simulations, with a $u-v$ coverage and thermal noise levels that are identical to those of the 2017 EHT observations of M87 (\citealt{PaperIII}; see Fig.~\ref{fig:test1_uv} for the $u-v$ coverage).

In the following subsection, we discuss a number of cases that correspond to different morphologies of the underlying images.

\begin{figure*}[t]
  \centerline{ 
   \includegraphics[width=0.33\textwidth]{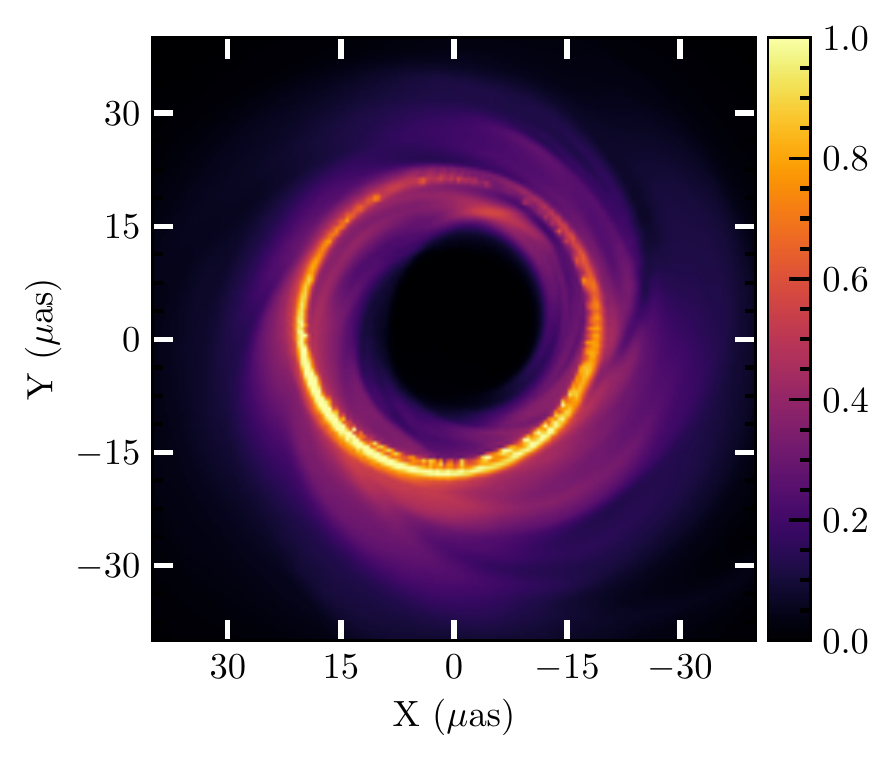}
   \includegraphics[width=0.33\textwidth]{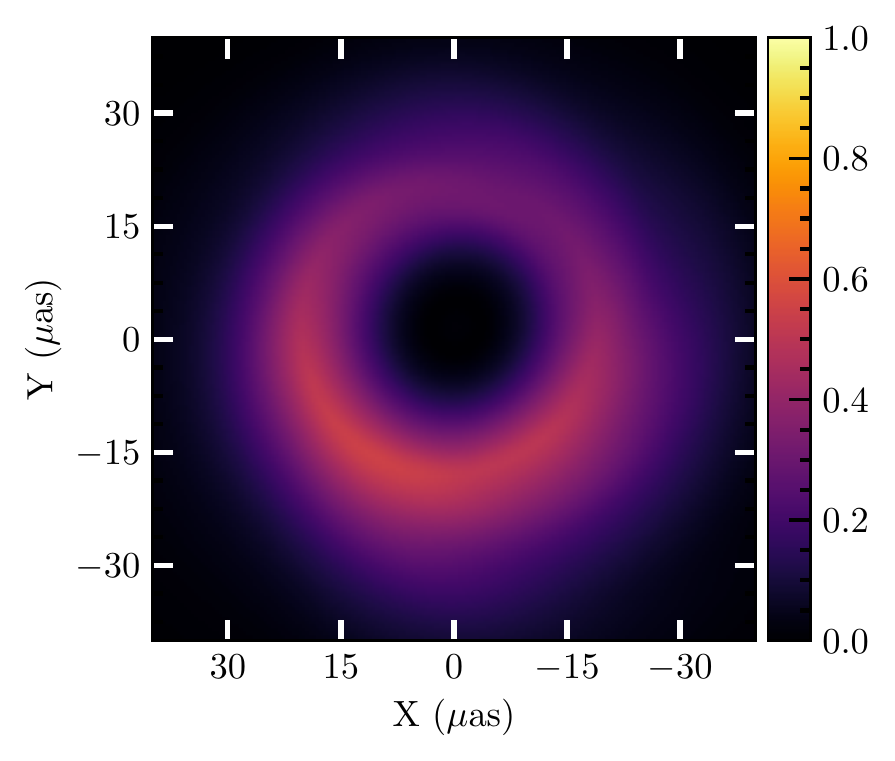}}
   \centerline{
\includegraphics[width=0.33\textwidth]{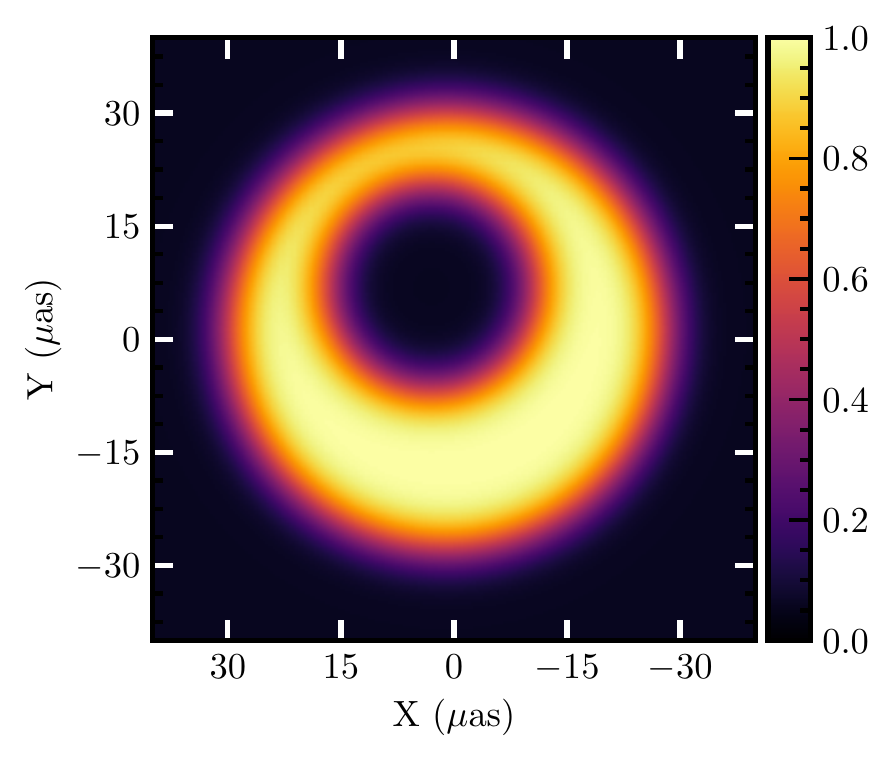}
\includegraphics[width=0.33\textwidth]{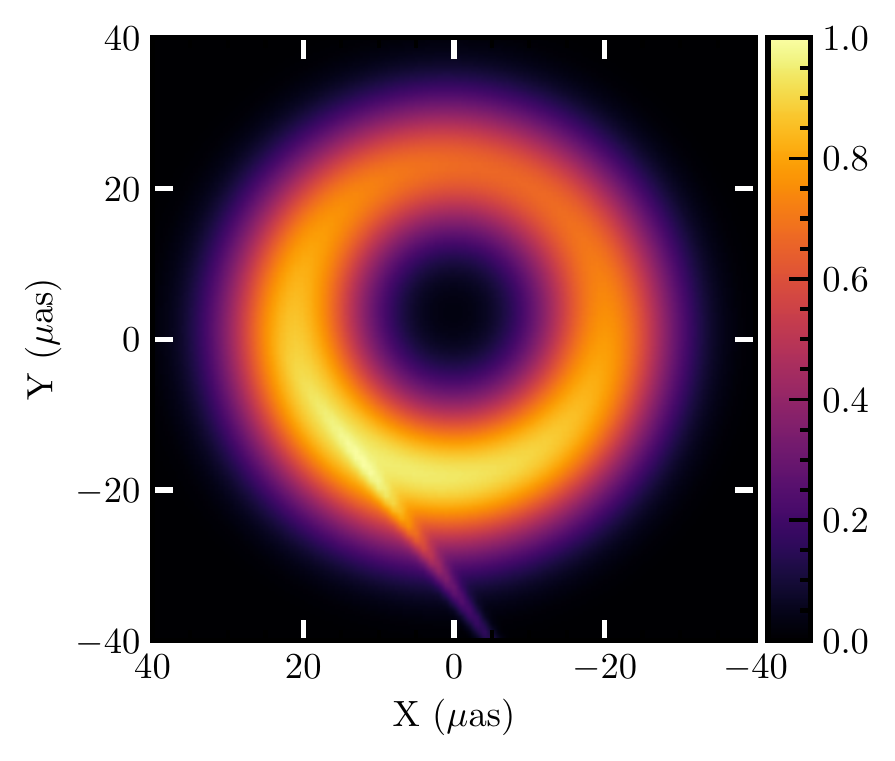}}
    \caption{{\em (Upper Left)\/} A second snapshot image from a MAD GRMHD+GR Radiative Transfer simulation with parameters that are relevant to the M87 black hole. {\em (Upper Right)\/} The same snapshot filtered with an $n=2$ Butterworth filter with a cutoff at 15~G$\lambda$. {\em (Lower Left)\/} The best-fit crescent image to the synthetic EHT data generated from the snapshot on the upper left panel; the width and orientation of the crescent image does not match the ones of the image used to generate the synthetic data. {\em (Lower Right)\/} The best-fit crescent image, when an additional elliptical Gaussian component is added to the model. The addition of this component leads to a more reasonable crescent width and an orientation that resembles that of the original image. However, the width and orientation of the Gaussian component does not match any particular structure in the original image and its minor axis is lined up with one of the prominent orientations of EHT baselines.}
        \label{fig:MAD_47_image} 
\end{figure*}

\subsection{Images with Dominant Crescent Structures}

The low inferred inclination of the black-hole spin in M87 substantially reduces the brightness asymmetry that is caused by the Doppler effect (compare, e.g., with the case of high-inclination simulations for Sgr~A* reported in \citealt{Chan2015}). As a result, if the emission is highly localized near the black-hole horizon, the resulting image will be dominated by a crescent structure with mild asymmetry. An example of such a snapshot from our MAD simulation is shown in the left panel of Figure~\ref{fig:MAD_134_image}. The North-South brightness asymmetry in this snapshot is 2:1, similar to what is inferred in the M87 image. The middle panel of the same figure shows the same snapshot but filtered with a cutoff baseline of 15~G$\lambda$ to suppress structures that are not accessible to the EHT array; for this and the following figures, this is achieved by using an $n=2$ Butterworth filter (see Psaltis et al.\ 2020).

\begin{figure*}[t]
  \centerline{  
  \includegraphics[width=0.4\textwidth]{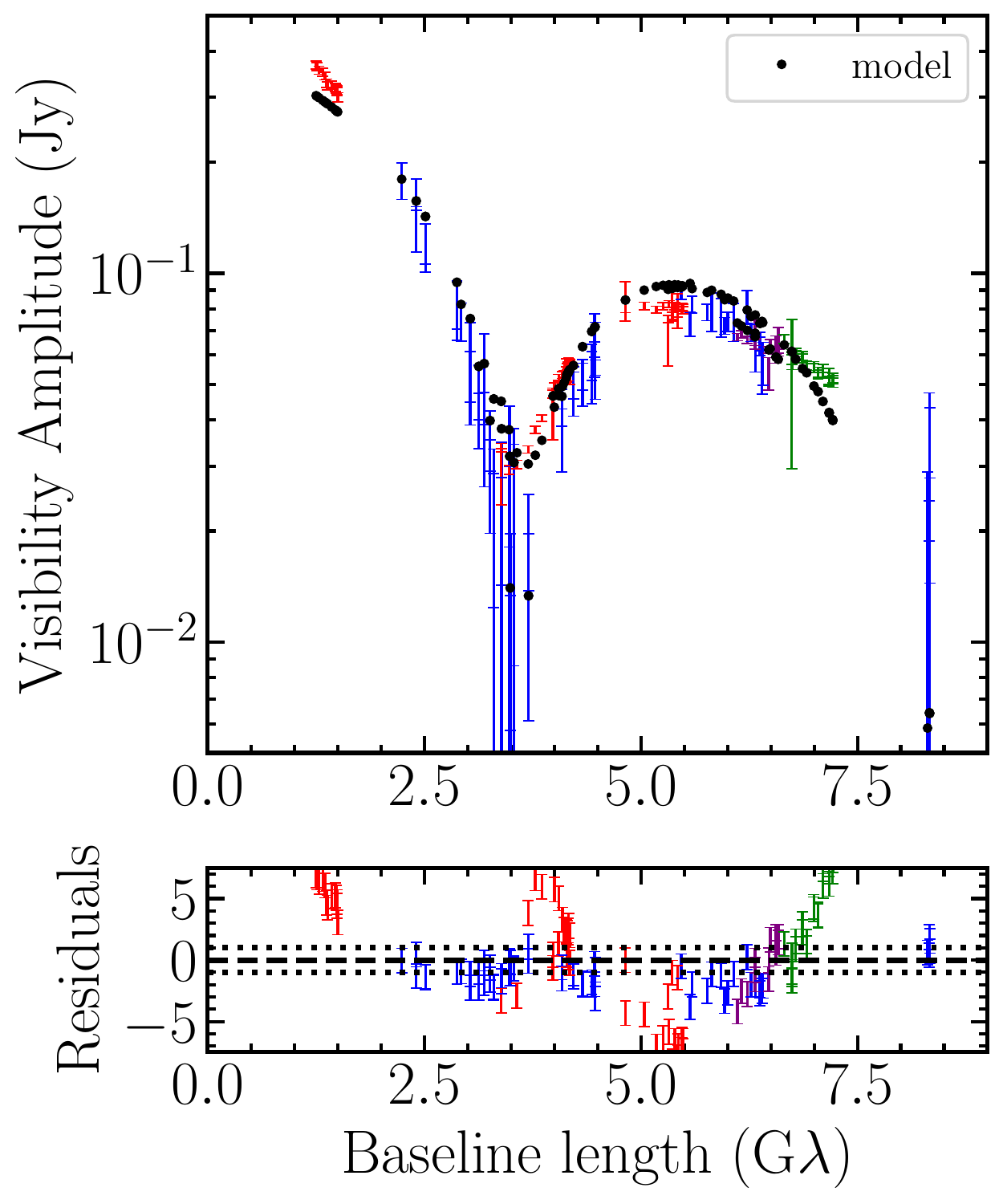}
  \includegraphics[width=0.4\textwidth]{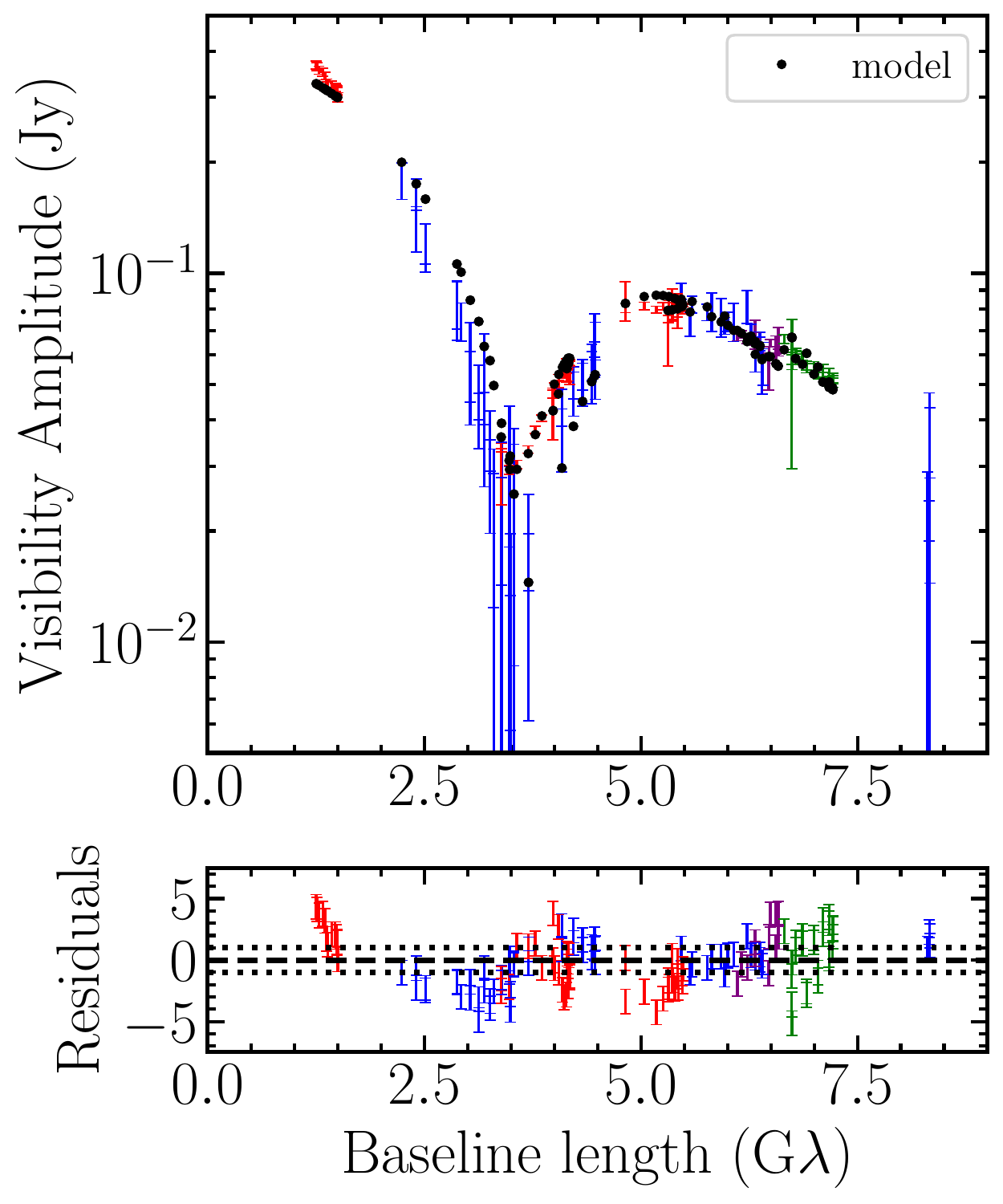}
  }
    \caption{Same as Fig.~\ref{fig:MAD_134_visamp} but for the synthetic data from the snapshot shown in the upper left panel Fig.~\ref{fig:MAD_47_image}. {\em (Left)\/} Fit to the synthetic data with a simple broadened crescent model; substantial residuals exist, especially for the red and green baselines; these residuals are the reason behind the biased width and orientation of the best-fit image (see bottom left panel of Fig.~\ref{fig:MAD_47_image}). {\em (Right)\/} Fit to the same synthetic data but with a broadened crescent model to which an asymmetric Gaussian component was added; the addition of this component substantial reduces the residuals.
     \label{fig:MAD_47_visamp}}
\end{figure*}

Figure~\ref{fig:MAD_134_visamp} shows the synthetic data for visibility amplitudes and closure phases generated from this image.  For the assumed parameters of the black hole, the radius of the shadow is equal to 18.7~$\mu$as and, for an image dominated by an infinitesimal bright ring surrounding the shadow, the first minimum in the visibility amplitude is expected at a baseline length of $\simeq 4.2$~G$\lambda$. The synthetic data show instead a visibility amplitude minimum at a baseline length of $\simeq 3.8$~G$\lambda$, which corresponds to a radius of the emitting region of $\simeq 20.5$~$\mu$as, signifying a broader emission region.

Figure~\ref{fig:MAD_134_visamp} also shows the best-fit crescent model that we obtained after running \march\ for 50 million steps and 40 replica temperatures. For the purposes of this test, we allowed for the crescent image to have softer edges by multiplying the complex visibilities of the model (eq.~[\ref{eq:ring}]) with a Gaussian function, which is equivalent to convolving the crescent image by the corresponding Gaussian kernel. The right panel of Figure~\ref{fig:MAD_134_image} shows the image that corresponds to the best-fit model parameters. Clearly, when the underlying image is dominated by a crescent structure, then fitting a simple crescent model to synthetic EHT data results in only small residuals and a model image that closely matches the properties of the one used to generate the synthetic data.

Figure~\ref{fig:MAD_134_corner} shows the posteriors of the model parameters. The very high signal-to-noise ratio of the data, which exceeds 100 for some of the data points, together with the small number of model parameters, leads to highly constrained values for the latter. Large correlations can be seen primarily between two pairs of model parameters. The posteriors in the fractional width $\psi$ and the radius $R$ of the crescent are correlated according to equation~(\ref{eq:Rcor}), as their values are primarily determined by the baseline length of the first minimum in the visibility amplitude\footnote{The definition of the fractional width as $\psi \sim 1/R$ would introduce an anticorrelation but this explains neither the shape nor the range of values of the correlated posteriors.}. At the same time, the posteriors in the FWHM of the broadening Gaussian kernel and the fractional width of the crescent are anticorrelated because the same effective broadening of the crescent can be achieved by either increasing the fractional width of the crescent or by broadening it by a wider Gaussian kernel.

Figure~\ref{fig:MAD_134_influence} shows the relative influence that each of the synthetic data points with realistic uncertainties exerts on the inference of the model parameters, calculated using equation~(\ref{eq:influence}). In order to separate the two types of influential data points (i.e., those with intrinsically large leverage versus the outliers that are not well described by the model), we have used filled and open circles to denote the data points for which the difference between the measurement and the model prediction is less than or more than 2.5 times the measurement uncertainty, respectively. We have also color-coded the data points as in Figure~\ref{fig:test1_uv} to show the orientations of the various baselines. As expected from the discussion in \S4, the data points near the minimum of the visibility amplitude at $\sim 4$~G$\lambda$ are described well by the model and have high leverage in determining the model parameters. In contrast, a number of data points exist at various baselines that are not described well by the model (i.e., they have large residuals) but exert substantial influence in determining the model parameters because they are outliers. In this case of a realistic image, which is not described adequately by a simple crescent model, and of heteroscedastic data, the posteriors of the model parameters are determined by a combination of the influence of data points of high leverage and of the outliers.

The fact that an image appears to be dominated by a crescent structure does not mean that the resulting synthetic data will always be well described by a crescent model. This is demonstrated in Figure~\ref{fig:MAD_47_image}, which shows a snapshot with what appears to be a dominant crescent structure. However, fitting the resulting synthetic data with a crescent model results in substantial residuals (see left panel of Fig.~\ref{fig:MAD_47_visamp}) and the corresponding best-fit image does not match the width or the orientation of the original image (see lower left panel of Fig.~\ref{fig:MAD_47_image}). 

It is tempting to try to improve the quality of the fit by adding elliptical Gaussian components to the crescent model. Adding even a single such elliptical Gaussian increases the number of model parameters from 6 (see Fig.~\ref{fig:MAD_134_corner}) to 12, with the additional six parameters being the centroid positions along the two axes of the Gaussian component, its major and minor axes, its position angle, and its flux. This large increase in the number of model parameters indeed leads to substantially smaller residuals, as can be seen in the right panel of Figure~\ref{fig:MAD_47_visamp}. 

Comparing the best-fit image to the original snapshot from the simulation (see lower right panel of Fig.~\ref{fig:MAD_47_image}) reveals that the additional Gaussian component seems to be adding flux at the lower-left quadrant of the image that has additional structure. However, it is extremely narrow and its minor axis in the image domain (i.e., the major axis in the visibility domain) is aligned with the orientation of the green baselines. This is not surprising, given that the main utility of the additional elliptical Gaussian component in this particular example has been to reduce the residuals along the red and green color-coded baselines, which have the highest signal-to-noise ratios, without affecting substantially the blue color-coded ones, which have lower signal-to-noise ratios and were well described by the simple crescent model (see Fig.~\ref{fig:test1_uv} for the color code of the baselines). The size, asymmetry, and orientation of the additional Gaussian component is simply an artifact of the sparse coverage of the EHT array and the orientations of its baselines.

Nevertheless, even though the properties of the additional Gaussian component are not reliable, its presence allowed the best-fit crescent model to have a width and an orientation that more closely matches that of the original snapshot compared to the best-fit model without this extra component. In both examples described in this subsection, the dominant crescent shapes in the images caused deep minima in the visibility amplitudes at baselines $\simeq 3.8-4$~G$\lambda$ and, for this reason, the inferred sizes of the crescents are comparable to the true values.

\begin{figure*}[t]
  \centerline{ 
   \includegraphics[width=0.33\textwidth]{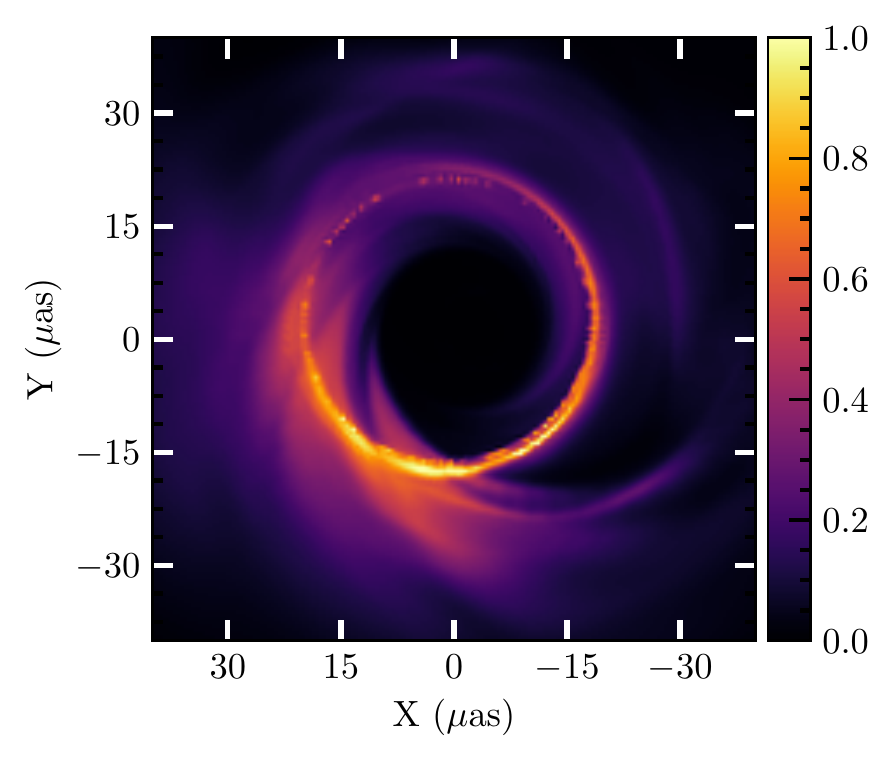}
   \includegraphics[width=0.33\textwidth]{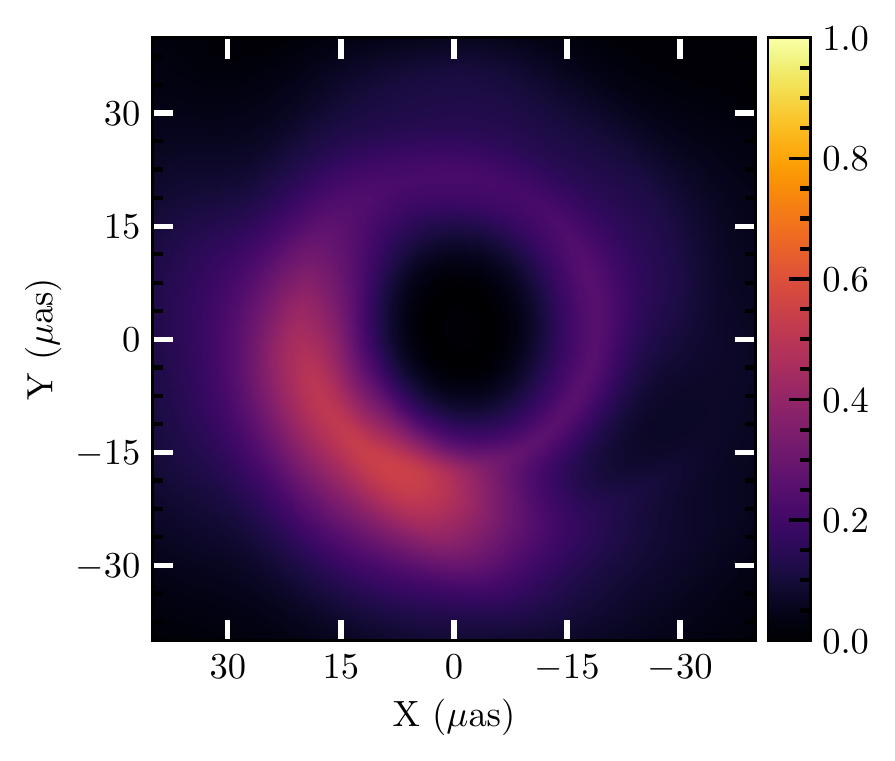}}
   \centerline{
\includegraphics[width=0.33\textwidth]{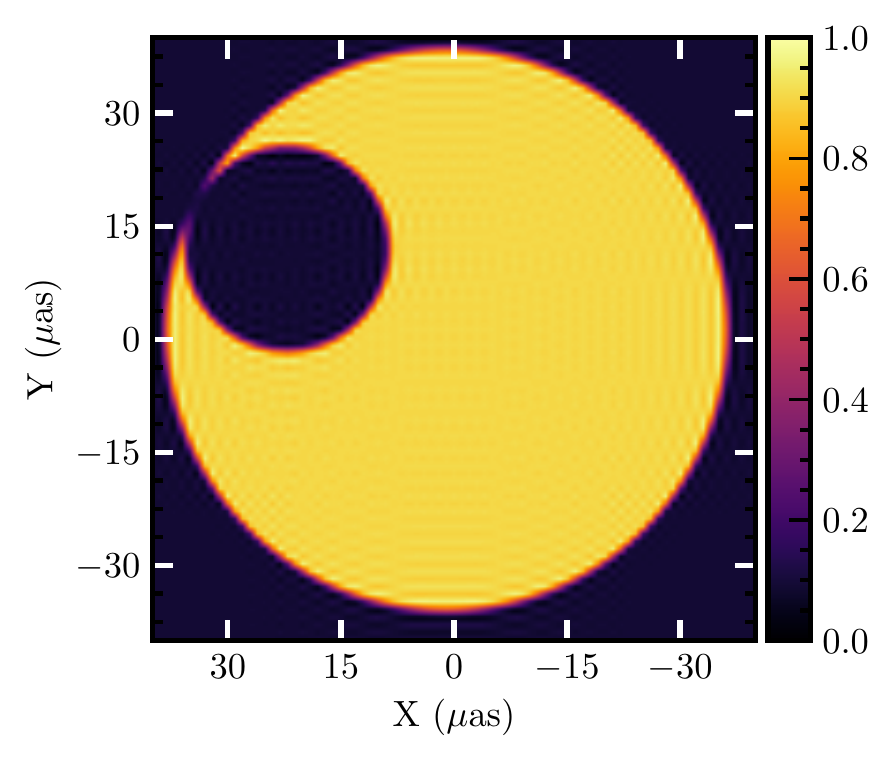}
\includegraphics[width=0.33\textwidth]{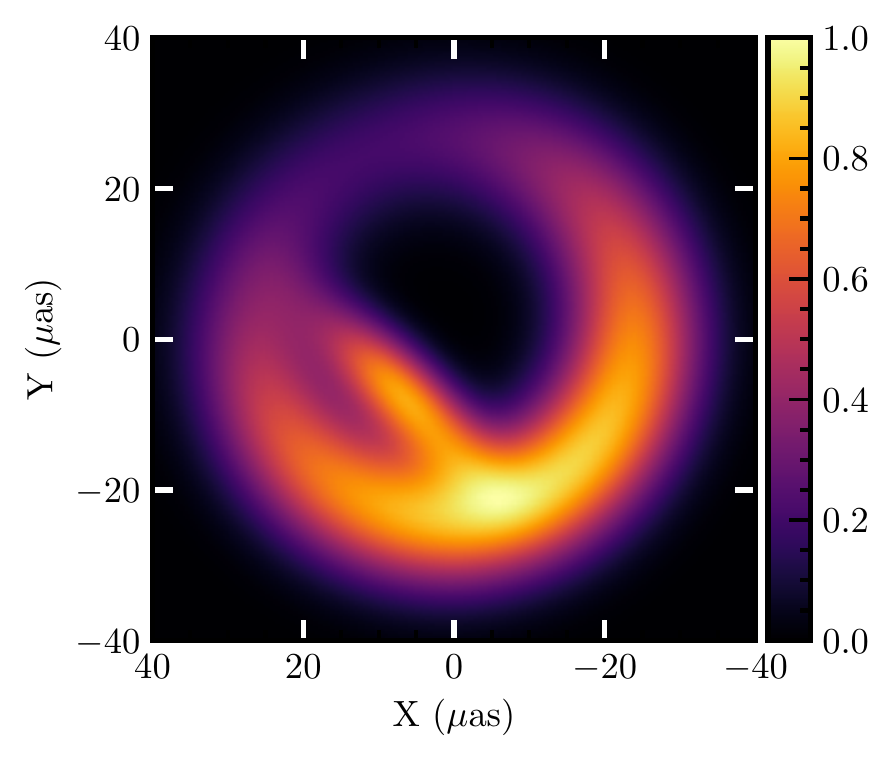}}
    \caption{{\em (Upper Left)\/} A third snapshot image from a MAD GRMHD+GR Radiative Transfer simulation with parameters that are relevant to the M87 black hole. This snapshot has substantial structure beyond the crescent shape {\em (Upper Right)\/} The same snapshot filtered with an $n=2$ Butterworth filter with a cutoff at 15~G$\lambda$. {\em (Lower Left)\/} The best-fit crescent image to the synthetic EHT data generated from the snapshot on the upper left panel; the width and orientation of the crescent image does not match the one used to generate the synthetic data. {\em (Lower Right)\/} The best-fit crescent image, when an additional elliptical Gaussian component is added to the model. The addition of this components leads to a crescent width (but not size or orientation) that resembles that of the original image. However, this additional component does not match in detail any particular structure in the original image and, as in Fig.~\ref{fig:MAD_47_image},  its minor axis is again lined up with one of the prominent orientations of EHT baselines. }
        \label{fig:MAD_191_image} 
\end{figure*}

\begin{figure*}[t]
  \centerline{  
  \includegraphics[width=0.4\textwidth]{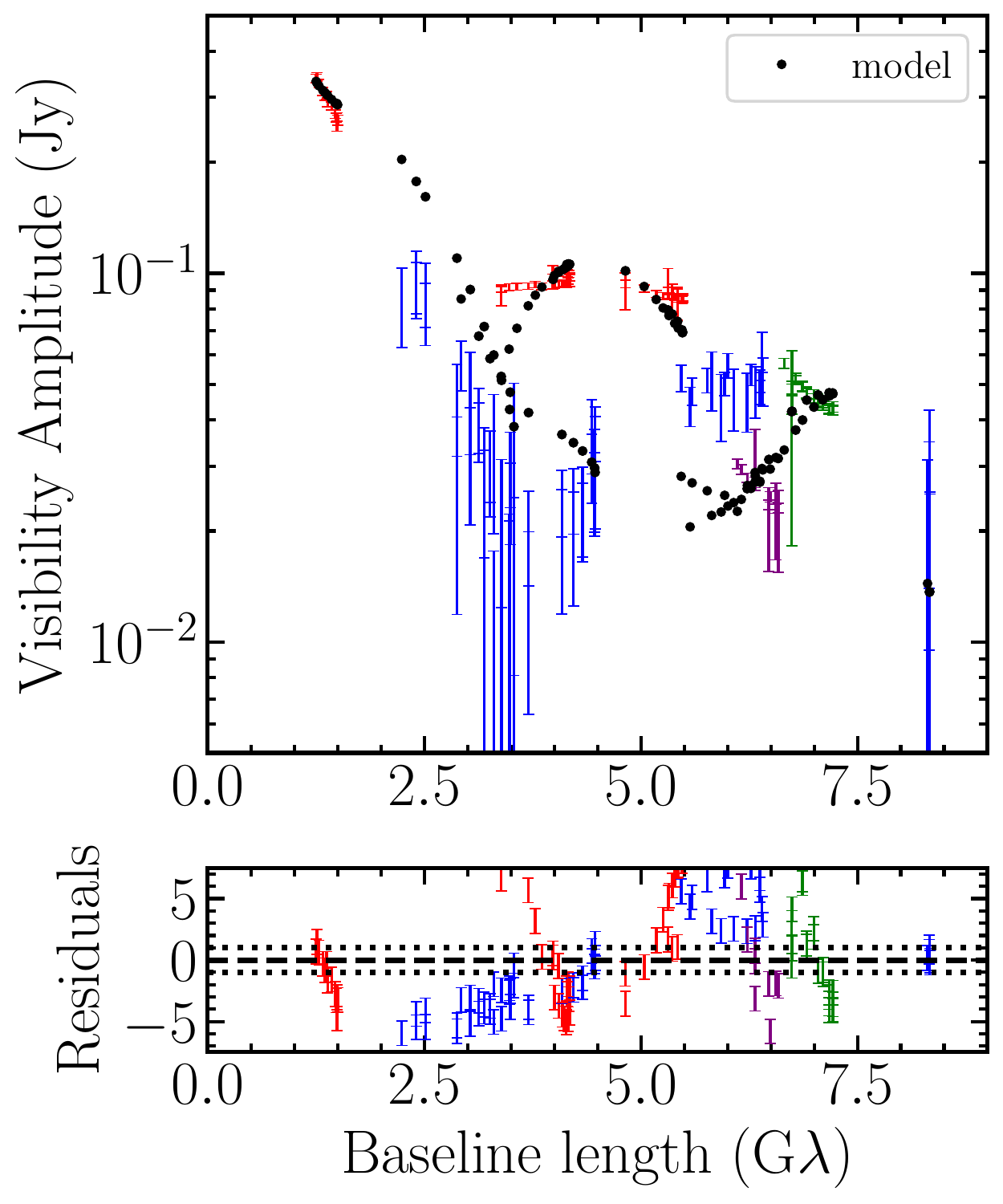}
  \includegraphics[width=0.4\textwidth]{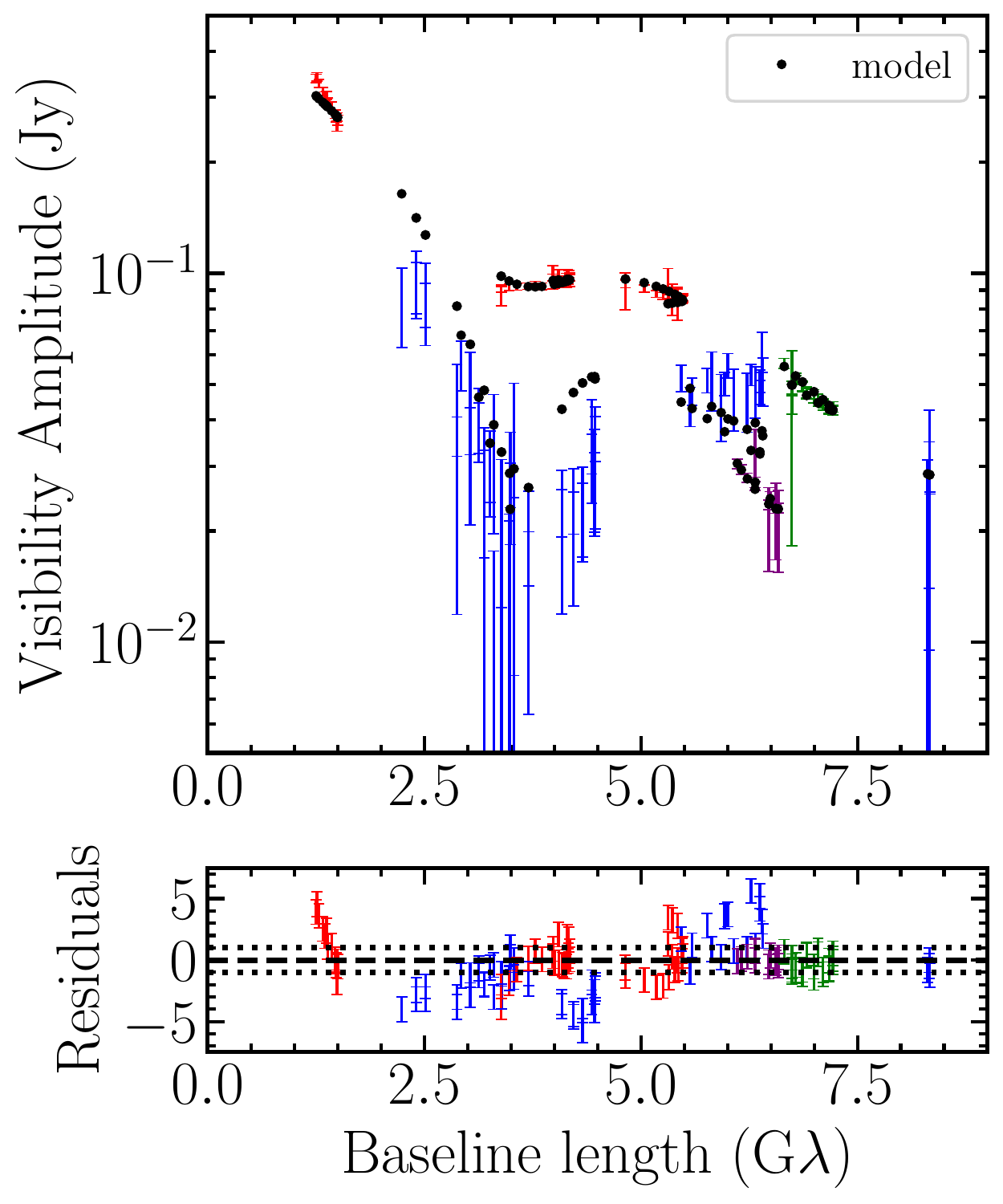}
  }
    \caption{Same as Fig.~\ref{fig:MAD_134_visamp} but for the synthetic data from the snapshot shown in the upper left panel Fig.~\ref{fig:MAD_191_image}. {\em (Left)\/} Fit to the synthetic data with a simple broadened crescent model; substantial residuals exist, especially for the red and green baselines; these residuals are the reason behind the biased width and orientation of the best-fit image (see bottom left panel of Fig.~\ref{fig:MAD_47_image}). {\em (Right)\/} Fit to the same synthetic data but with a broadened crescent model to which an asymmetric Gaussian component was added; the addition of this component substantial reduces the residuals.
     \label{fig:MAD_191_visamp}}
\end{figure*}

\subsection{Images with Sub-Dominant Crescent Structures}

Figure~\ref{fig:MAD_191_image} shows a third example of a snapshot from a MAD simulation with characteristics that are frequently encountered. In this example, there is substantial structure beyond the simple crescent shape that persists even after we filtered the image with the Butterworth filter in order to suppress visually the structures that are not accessible to the EHT data. 

The synthetic data from this snapshot (Fig.~\ref{fig:MAD_191_visamp}) indeed do not show the characteristic ringing structure with pronounced deep minima and the overlapping secondary maxima of the crescent shapes along all orientations. It is, therefore, expected that fitting a simple broadened crescent model to the synthetic data (left panel of Fig.~\ref{fig:MAD_191_visamp}) leaves substantial residuals at all baselines. The resulting best-fit image (lower-left panel of Fig.~\ref{fig:MAD_191_image}) does not match any of the structures in the image from which the synthetic data were generated. 

Adding an additional Gaussian component to the model substantially improves the quality of the fit and reduces the residuals, especially along the red- and green-baselines that have the data with the highest signal-to-noise ratios (right panel of Fig.~\ref{fig:MAD_191_visamp}). The best-fit image (lower right panel of Figure~\ref{fig:MAD_191_image}), however,  shows structures that again do not match those of the image from which the synthetic data were generated. The width of the crescent shape is reasonable but neither its orientation nor its size are correct. Moreover, as in Fig.~\ref{fig:MAD_47_image} in the last example discussed in the previous subsection, the minor axis of the additional Gaussian component in the image domain (i.e., the major axis in the visibility domain) is aligned with the orientation of the green baselines. 

\section{Conclusions}

In this paper, we introduced \march, a new Markov Chain Monte Carlo algorithm with parallel tempering for fitting theoretical models to the interferometric data from the Event Horizon Telescope. We then used this algorithm to explore the biases in the inferred parameters of crescent models that we fit to synthetic EHT data based on images from GRMHD simulations. We explored, in particular, biases that are introduced by the uneven influence of data points, the heteroscedastic nature of the EHT data, and the sparse coverage of the interferometric $u-v$ plane with the EHT baselines. 

When the images are dominated by strong crescent structures, we find that there are a number of salient features in the interferometric data, such as a deep minimum in the visibility amplitude at a baseline length determined primarily by the size of the image. Data obtained in the vicinity of such salient features exert a disproportionate influence on the values of the inferred model parameters, artificially reducing the widths of their posteriors. 

When the underlying images have structures that are not well characterized by crescent shapes, fitting crescent models to synthetic EHT data reveals substantial residuals, as expected. Attempting to ameliorate these residuals by adding compact (e.g., Gaussian) components to the models may improve the quality of the fit and reduce the residuals. However, the properties of the additional components are strongly biased by the locations and orientations of the EHT baselines and are not trustworthy. Such Gaussian components with similarly large aspect ratios and orientations were deemed necessary to fit the actual M87 data observed in 2017 with the EHT (see Fig.~5 of \citealt{PaperVI}).

When the underlying image has a high degree of symmetry, the sparse baseline coverage of the EHT array often biases even the inferred overall orientation of the underlying image to align with the orientations of the primary baselines (see, e.g., Fig.~\ref{fig:test1_angle}). If we were to match the features of a crescent model with those of a snapshot from a GRMHD simulation, we would identify the "orientation" of the crescent image (which measures the orientation of displacement between the outermost, positive disk and the innermost, negative one) with the perpendicular orientation to that of the black-hole spin. Figure~\ref{fig:test1_angle} shows that the orientations of a crescent model that was fit to synthetic EHT data of a symmetric structure appear to get biased towards a position angle of $\simeq 150^\circ$ EofN. This would imply a bias in the inferred orientation of the black-hole spin towards an orientation $\simeq 150+90=240^\circ$ EofN, which is very similar to the value with the highest posterior inferred from fitting crescent models to the 2017 EHT data and casts doubt on that inference (see, e.g., Figure~9 of \citealt{PaperV}). 

In all cases, the robust detection of salient features leads to the determination of the characteristic sizes of crescent images that are within a small fractional distance from (albeit often statistically inconsistent with) the true sizes of the underlying images. When such salient features are detected, as was the case with the 2017 EHT data on M87, the size of the black-hole shadow and, hence, of the black-hole mass is not significantly biased. In order to obtain accurate measurements of additional aspects of the image, such as its orientation, asymmetry, and presence of additional structures, requires a more complete coverage of the $u-v$ space with the addition of more stations in the EHT array. A number of such stations have been introduced since the 2017 observations, such as the GLT telescope in Greenland, the 12m telescope on Kitt Peak, and NOEMA in the French Alps, which will help substantially in alleviating the biases introduced by the sparse baseline coverage of the current EHT.

\acknowledgements

We thank K.\ Satapathy and T.\ Trent for useful discussions. This work was supported in part by NSF PIRE grant 1743747 and NSF grant AST 1715061. L.\,M.\ acknowledges support from an NSF Astronomy and Astrophysics Postdoctoral Fellowship under award no.\ AST-1903847.  C.\,R.\ acknowledges support from NSF Graduate Research Fellowship Program Grant DGE-1746060. All ray tracing calculations were performed with the El Gato GPU cluster at the University of Arizona that is funded by NSF award 1228509.

\bibliographystyle{apj}

\bibliography{mcmc.bib}

\end{document}